\documentclass[floatfix,aps,prd,showpacs,amsmath,nofootinbib,preprintnumbers,twocolumn]
{revtex4}
\usepackage{graphicx}
\usepackage{colordvi}

\newcommand{\tk}{\tilde{k}}
\def\be{\begin{equation}}
\def\ee{\end{equation}}
\def\ba{\begin{eqnarray}}
\def\ea{\end{eqnarray}}

\begin{document}

\title{The Impact of  
Stochastic Primordial Magnetic Fields on the Scalar Contribution to 
Cosmic Microwave Background Anisotropies}

\author{Fabio Finelli}\email{finelli@iasfbo.inaf.it}
\affiliation{INAF/IASF-BO,
Istituto di Astrofisica Spaziale e Fisica
Cosmica di Bologna \\
via Gobetti 101, I-40129 Bologna - Italy}
\affiliation{INAF/OAB, Osservatorio Astronomico di Bologna,
via Ranzani 1, I-40127 Bologna -
Italy}
\affiliation{INFN, Sezione di Bologna,
Via Irnerio 46, I-40126 Bologna, Italy}
\author{Francesco Paci}\email{paci@iasfbo.inaf.it}
\affiliation{Dipartimento di Astronomia,
Universit\`a degli Studi di Bologna,
\\ via Ranzani, 1 -- I-40127 Bologna -- Italy}
\affiliation{INAF/IASF-BO,
Istituto di Astrofisica Spaziale e Fisica
Cosmica di Bologna \\
via Gobetti 101, I-40129 Bologna - Italy}
\affiliation{INFN, Sezione di Bologna,
Via Irnerio 46, I-40126 Bologna, Italy}
\author{Daniela Paoletti}\email{paoletti@iasfbo.inaf.it}
\affiliation{Dipartimento di Fisica,
Universit\`a degli Studi di Ferrara,
\\ via Saragat, 1 -- I-44100 Ferrara -- Italy}
\affiliation{INAF/IASF-BO,
Istituto di Astrofisica Spaziale e Fisica
Cosmica di Bologna \\
via Gobetti 101, I-40129 Bologna - Italy}
\affiliation{INFN, Sezione di Bologna,
Via Irnerio 46, I-40126 Bologna, Italy}

\begin{abstract}
We study the impact of a stochastic background of primordial magnetic 
fields on the scalar contribution of CMB anisotropies and on the matter 
power spectrum. 
We give the correct initial conditions for cosmological 
perturbations and the exact expressions for the energy density and Lorentz 
force associated to the stochastic background of primordial magnetic fields, 
given a power-law for their spectra cut at a damping scale. 
The dependence of the CMB temperature and polarization spectra on the relevant 
parameters of the primordial magnetic fields is illustrated.
\end{abstract}

\maketitle

\section{Introduction}

Large scale magnetic fields are almost everywhere in the universe, 
from galaxies up to those present in galaxy clusters and 
in the intercluster medium \cite{GR}. 
The origin of these magnetic fields depends on the size 
of the objects and may become mysterious for the largest ones.
The dynamo mechanism provides a 
mechanism to explain the observed magnetic field associated to galaxies, 
whereas those associated to clusters may be generated 
by gravitational compression starting from an initial seed.

The requirement of an initial seed for magnetic fields observed in 
galaxies and galaxy clusters leads directly to question the existence of 
primordial magnetic fields in the early universe. 
Cosmology described 
by an homogeneous and isotropic expanding metric neither supports 
a uniform magnetic field nor a gravitational amplification of gauge 
fields because of conformal invariance; 
the generation of large scale magnetic fields 
has therefore generated a lot of interest. 
A stochastic background (SB) of primordial magnetic fields (PMF) 
can provide the initial seeds for the large-scale magnetic fields observed 
and can leave imprints on different observables, as the CMB pattern 
of temperature and polarization anisotropies 
\cite{giovannini_review,subra_bologna} and the matter power spectrum.

A SB of PMF carries zero energy and pressure 
at homogeneous level in a Robertson-Walker metric. 
It carries however perturbations, of any kind, i.e. 
scalar, vector and tensor, and it is usually studied 
in a quasi-linear approximation, i.e. its EMT - quadratic in the magnetic 
field amplitude - is considered at the same footing as first order terms in 
a perturbative series expansion.
Vector \cite{MKK,vector} and tensor \cite{MKK,DFK,CDK} 
metric perturbations sourced by a PMF SB have been object of 
several investigations; beyond the technical simplicity of vector and tensor 
over scalar, a perfect fluid cannot support 
vector and tensor perturbations at linear order and therefore represent a 
key prediction of a PMF SB. We know however that temperature and 
polarization anisotropies 
sourced by scalar fluctuations with adiabatic initial conditions 
are a good fit to the whole set of observations; 
it is therefore crucial to investigate how a PMF SB 
can modify these scalar fluctuations. 
Analytic \cite{KR} and numerical \cite{KL,yamazaki,giovanninikunze} works in 
this direction have already been made. 
However a detailed analysis which takes into account the Lorentz 
force on baryons, a careful treatment of initial conditions and an 
accurate treatment of the Fourier spectra of PMF energy-momentum tensor 
is still lacking. 
As is clear in the following, our work address 
carefully both these issues. 

The goal of this paper is to investigate the impact 
of a stochastic background (SB) of primordial magnetic fields (PMF) on 
scalar cosmological perturbations and in particular on CMB temperature 
anisotropies and matter power spectrum. 
Our paper is organized as follows. In Section II we review how to add a 
fully inhomogenous SB of PMF treated in the one-fluid plasma description 
\cite{giovannini_review} to the Einstein-Boltzmann system of equations. 
In Sections III and IV we review the baryons evolution 
and we give the initial conditions 
for cosmological perturbations in a form suitable to be plugged in 
most of the Einstein-Botzmann codes. In Section V we give the 
PMF energy density and Lorentz force power spectra and compare our results 
with the ones given in the literature. In Sections VI-VIII we show 
the results obtained by our modification of the Einstein-Boltzmann 
code CAMB \cite{CAMB}
for cosmological scalar perturbations, CMB spectrum of temperature 
and polarization, 
matter power spectrum, respectively.
In the Appendix we show the detailed calculations for the convolution 
integrals leading to the energy density and Lorentz force, starting from 
a power-law spectrum sharply cut at a given scale for the PMF.

\section{Stochastic Magnetic Fields and Cosmological Scalar Perturbations}

We model a SB of PMFs as a fully inhomogenous component, considering 
$B^2$ at the same level of metric and density fluctuations in a 
perturbative expansion \footnote{Note that in such a way we do not take 
into account the modification of the sound speed of baryons induced by 
PMFs, pionereed in \cite{ADGR}, since it would be 
technically of second order in the equations of motion. However, since 
the baryons speed of sound goes rapidly to zero in the matter dominated era, 
this effect, leading to a shift in the Doppler peaks, may be anyway 
important.}.  
Although a SB of PMFs carries no energy at the homogeneous level, it 
affects scalar cosmological perturbations in three different ways. 
First, inhomogeneous PMFs carry energy density and pressure and therefore 
gravitate at the level of perturbations.
Second, inhomogeneous PMFs have anisotropic stress - differently from 
perfect fluids - which adds to the photon and 
neutrino ones, with the caveat that the photon anisotropic stress is 
negligible before the decoupling epoch.
Last, but not least, the induced Lorentz force acting on baryons, 
affects also photons during the 
tight coupling regime.

Since the EMT of PMF at homogeneous level is zero,  
at linear order PMFs evolve like a stiff source and therefore it 
is possible to discard all the back reactions of the fluid or gravity 
onto the SB of PMF.
Before the decoupling epoch the electric conductivity of the 
primordial plasma is very large,
therefore it is possible at the first order to consider the 
infinite conductivity limit.
In this limit the induced electric field is zero. 
Within the infinite conductivity limit the SB of 
PMF time evolution simply reduces to :
${\bf B}({\bf x},\tau)={\bf B}({\bf x})/a(\tau)^2$.
\footnote{We choose the standard convention in which at present time $t_0$, 
$a(t_0) = 1$.}

The evolution of the metric perturbations in the presence of PMF is
governed by the Einstein equations:
\be
G_{\mu\nu}=8\pi (T_{\mu\nu}+\tau_{\mu\nu}^{\rm PMF})\,,
\ee
In the approximation in which the induced electric field is vanishing
(i.e. the infinite conductivity limit)
the energy momentum tensor of the electromagnetic field becomes:
\begin{eqnarray}
\tau^{0 \, {\rm PMF}}_0 &=& - \rho_B = 
- \frac{|{\bf B} ({\bf x})|^2}{8\pi  a^4} \,, \label{EMT}\\
\tau^{0 \, {\rm PMF}}_i &=& 0 \,, \\
\tau^{i \, {\rm PMF}}_j &=& \frac{1}{4\pi  a^4} \left(
\frac{|{\bf B} ({\bf x}) |^2}{2} \delta^i_j - 
B_j ({\bf x}) B^i ({\bf x}) \right) 
\label{EMT3}.
\end{eqnarray}
In the Fourier
space\footnote{As Fourier transform and its inverse, we use
- in agreement with \citep{MB} -:
\be
Y (\vec k, \tau)=\int \frac{d^3x}{(2 \pi)^3}
e^{- i \vec k \cdot \vec x} Y (\vec x, \tau)
\, ,\,\,\,\,\, Y (\vec x, \tau)=\int d^3 k
e^{i \vec k \cdot \vec x} Y (\vec k, \tau) \,.
\label{Fourier}
\ee
where $Y$ is a generic function.} the Einstein equations with 
the contribution of PMF in the synchronous gauge are:
\begin{eqnarray}
k^2 \eta-\frac{1}{2} {\mathcal{H}} \dot h &=& 4\pi
G a^2 (\Sigma_n\,\rho_n\delta_n+\rho_B)  \,,\nonumber \\
k^2 \dot\eta &=& 4\pi G a^2 \Sigma_n(\rho_n+P_n)\theta_n \,, \nonumber\\
\ddot h +2 {\mathcal{H}} \dot h -2 k^2 \eta &=& - 8\pi G a^2
(\Sigma_n \,c_{s \, n}^2\rho_n\delta_n  \nonumber \\
& &
+\frac{\rho_B}{3}) \nonumber \,, \\
\ddot h+6\ddot \eta +2 {\mathcal{H}} (\dot h+6 \dot\eta)-2 k^2\eta
&=& -24 \pi G a^2 \times \nonumber \\
& & [\Sigma_n(\rho_n+P_n) \sigma_n+\sigma_B] ,
\label{Einsteineqs}
\end{eqnarray}
where by $n$ we mean the number of components, i.e. baryons, cold dark 
matter (CDM), photons and neutrinos.
The conservation of the PMF EMT  - $\nabla_\mu \tau^{\mu \, {\rm PMF}}_\nu=0$ - 
simply reduces to :
\be
\sigma_B=\frac{\rho_B}{3}+L \,,
\label{stress_B}
\ee
where $\sigma_B$ represents the PMFs anisotropic stress and $L$ the 
Lorentz force. The energy density of PMF evolves like radiation:
$\rho_B({\bf x},\tau)=\rho_B({\bf x},\tau_0)/a(\tau)^4$.

\section{Baryons Evolution}

The presence of PMFs in a plasma which contains charged particles induces
a Lorentz force on these particles, that, in the primordial plasma, are baryons.
The general expression for the Lorentz force is \citep{KR}:
\be
L_i(x,\tau_0) = \frac{1}{4 \pi}\left[ B_j( {\bf x} )\nabla_j B_i( {\bf x} )-
\frac{1}{2} \nabla_i B^2({\bf x}) \right] \, ,
\ee
where ${\bf L}(x,\tau)=\frac{{\bf L}(x,\tau_0)}{a^4}$.\\
We are interested only in the scalar perturbations and 
the scalar part of the Lorentz force defined as 
$\nabla^2L^{(S)} \equiv \nabla_i L_i$ is therefore:
\be
\nabla^2L^{(S)} = \frac{1}{4 \pi}
\Big[(\nabla_iB_j( {\bf x} ))\nabla_j B_i( {\bf x} )-\frac{1}{2} 
\nabla^2 B^2 ({\bf x}) \Big] \, .
\label{LFS}
\ee

In the presence of an electromagnetic source the conservation equations 
of the baryon component of the primordial fluid becomes:
\ba
\nabla_\mu \delta T^{\mu\nu \, {\rm baryons}} \propto 
F^{\mu\nu}J_\mu
\ea
where $J_\mu$ is the quadrivector of the density current and $F^{\mu\nu}$ 
is the Maxwell tensor.
The primordial plasma can be considered globally neutral, 
this leads to $J_0=0$ and therefore to the fact that the energy conservation 
of baryons is not modified by the presence of the Lorentz term.
The Euler equation for baryons in instead affected by the Lorentz force and 
the scalar part is therefore \cite{giovannini_review}:
\be
\dot \theta_b = -{\mathcal{H}} \theta_b+k^2 c_{sb}^2 \delta_b
-k^2\frac{L}{\rho_b} \,.
\label{baryonseq}
\ee
Now we study how the tight-coupling regime is modified by the presence 
of a SB of PMF \cite{giovannini_tg}.
The Euler equation for photons during the tight-coupling regime is:
\be
\dot\theta_\gamma=k^2\Big(\frac{\delta_\gamma}{4} - \sigma_\gamma\Big)
+a n_e \sigma_T(\theta_b-\theta_\gamma)
\ee
Combining the photons and baryons equations gives:
\be
\dot \theta_b=\frac{-{\mathcal{H}}\theta_b+c_s^2 k^2\delta_b
+k^2R\Big(\frac{\delta_\gamma}{4}-\sigma_\gamma \Big)+
R(\dot\theta_\gamma-\dot\theta_b)-\frac{k^2L}{\rho_b}}{(1+R)}\nonumber\,,
\ee
with: 
\ba
\dot\theta_b-\dot\theta_\gamma &=& \frac{2R}{(1+R)}{\mathcal{H}}
(\theta_b-\theta_\gamma)+\frac{\tau}{(1+R)}
\left(-\frac{\ddot a}{a}\theta_b + \right. \nonumber \\
& & \left. -
\frac{{\mathcal{H}}k^2}{2}\delta_\gamma+k^2\Big(c_s^2 \dot \delta_b-
\frac{\dot\delta_\gamma}{4}\right)+{\mathcal{H}} k^2 \frac{L}{\rho_b}
\Big)\,,\nonumber
\ea

The photon Euler equation in tight coupling regime instead is:
\ba
\dot\theta_\gamma &=&-R^{-1}\Big(\dot\theta_b
+{\mathcal{H}}\theta_b-c_s^2k^2\delta_b+k^2
\frac{L}{\rho_b}\Big) \nonumber \\
& & +k^2\Big(\frac{\delta_\gamma}{4}-
\sigma_\gamma\Big)
\ea
We note that there is a term depending on the Lorentz force which 
disappears when the tight coupling ends, leaving the normal Euler equation for the photon velocity.

\section{Initial Conditions}

In order to study the effect of a PMF SB on scalar cosmological 
perturbations, the initial conditions for the latter deep 
in the radiation era are required (see \cite{giovannini_review} 
for the results in the longitudinal gauge). 
The magnetized adiabatic mode initial conditions in the synchronous 
gauge are given by \cite{tesi_daniela}:
\begin{eqnarray}
h&=&C_1(k\tau)^2 \nonumber \\
\eta&=&2C_1-\frac{5+4R_\nu}{6(15+4R_\nu)}C_1(k\tau)^2 + \nonumber \\
& & - \left[ \frac{\Omega_B(1-R_\nu)}{6(15+4R_\nu)}
+\frac{L_B}{2(15+4R_\nu)}\right]
(k\tau)^2\nonumber\\
\delta_\gamma&=&-\Omega_B-\frac{2}{3}C_1 (k\tau)^2+\left[ 
\frac{\Omega_B}{6}+\frac{L_B}{2(1-R_\nu)}\right] (k\tau)^2\nonumber\\
\delta_\nu&=&-\Omega_B-\frac{2}{3}C_1(k\tau)^2
-\left[ \frac{\Omega_B(1-R_\nu)}{6R_\nu}\frac{L_B}{2R_\nu}\right](k\tau)^2 
\nonumber\\
\delta_b&=&-\frac{3}{4}\Omega_B-\frac{C_1}{2}(k\tau)^2+\left[ 
\frac{\Omega_B}{8}+\frac{3L_B}{8(1-R_\nu)}\right] (k\tau)^2\nonumber\\
\delta_c&=&-\frac{C_1}{2}(k\tau)^2\nonumber\\
\theta_\gamma&=&-\frac{C_1}{18}k^4\tau^3+\left[ -\frac{\Omega_B}{4}
-\frac{3}{4}\frac{L_B}{(1-R_\nu)} \right] k^2\tau
\nonumber \\ 
& & +k \left[ \frac{\Omega_B}{72}+
\frac{L_B}{24(1-R_\nu)}\right] (k\tau)^3\nonumber\\
\theta_b&=&\theta_\gamma\nonumber\\
\theta_c&=&0\nonumber\\
\theta_\nu&=&-\frac{(23+4 R_\nu)}{18(15+4R_\nu)}C_1 k^4\tau^3+
\left[ \frac{\Omega_B(1-R_\nu)}{4 R_\nu}+\frac{3}{4}\frac{L_B}{R_\nu}\right] 
k^2\tau \nonumber\\
& &- \left[\frac{(1-R_\nu)(27+4R_\nu)\Omega_B}{72 R_\nu(15+4R_\nu)}+\frac{(27+4R_\nu)L_B}{24 R_\nu (15+4R_\nu)}\right] k^4\tau^3 \nonumber\\
\sigma_\nu&=&\frac{4 C_1}{3(15+4R_\nu)}(k\tau)^2-\frac{\Omega_B}{4R_\nu}
-\frac{3}{4}\frac{L_B}{R_\nu} \nonumber \\
& & + \left[ \frac{(1-R_\nu)}{R_\nu(15+4R_\nu)}
\frac{\Omega_B}{2}+\frac{3}{2}\frac{L_B}{R_\nu(15+4R_\nu)}\right] (k\tau)^2 \,,
\label{initialconds}
\end{eqnarray}
where $R_\nu=\rho_\nu/(\rho_\nu+\rho_\gamma)$ and $C_1$ 
is the constant which characterize 
the regular growing adiabatic mode as given in \citep{MB}.
We have checked that the result reported in \cite{giovanninikunze} 
and ours \cite{tesi_daniela} agree.

Note how the presence of a SB of PMFs induces a new independent mode
in matter and metric perturbations, i.e. the fully magnetic mode. 
This new independent mode is the particular solution of the 
inhomogeneous system of the Einstein-Botzmann 
differential equations: the SB of PMF treated as a stiff source acts indeed 
as a force term in the system of linear differential equations. 
Whereas the sum of the fully magnetic mode with the curvature one can be with 
any correlation as for an isocurvature mode, the nature of the fully 
magnetic mode - and therefore its effect - is different: the isocurvature 
modes are solutions of the homogeneous system (in which all the 
species have both background and perturbations), whereas the 
fully magnetic one is the solution of the inhomogeneous system sourced 
by a fully inhomogeneous component.

It is interesting to note the magnetic contribution drops from the 
metric perturbation at leading order, although is actually larger than the 
adiabatic solution for photons, neutrinos and baryons 
(the latter being tightly coupled to photons deep in the radiation era). 
This is due to a 
compensation which nullifies the sum of the leading contributions (in 
the long-wavelength expansion) in the single species energy densities 
and therefore in the metric perturbations. 
A similar compensation exists for a network of topological defects, 
which does not carry a background energy-momentum tensor as the PMF SB 
studied here \footnote{Note however that a network of topological defects 
does not scale with radiation and 
interacts only gravitationally with the rest of matter, i.e. a 
Lorentz term is absent.}.

\section{Magnetic Field Power Spectra}

Power spectra for the amplitude and the EMT of SB of PMF have been 
subject of several investigation \cite{DFK,MKK,KL,KR}.
We shall work in the Fourier space according to Eq. (\ref{Fourier}).   
We shall consider PMFs with a power law power spectrum, 
which therefore are characterize by two parameters: an amplitude $A$ 
and a spectral index $n_B$.
PMFs are suppressed by radiation viscosity on small scales:
we approximate this damping by introducing an ultraviolet cut-off 
in the power spectrum at the (damping) scale $k_D$. 

The two-point correlation
function for a statistically homogeneous and isotropic field is
\ba
\langle\vec{B_i^*}(\vec{k})\vec{B_j}(\vec{k'})\rangle=
\delta^3({\vec{k}}-{\vec{k'}})
& & \left[
(\delta_{ij}-\hat{k}_i\hat{k}_j) \frac{P_B(k)}{2} + \right. \nonumber \\
& & \left.  \epsilon_{ijl}
\frac{k_l}{k} P_H (k) \right] \,,
\label{spectrum}
\ea
where $\epsilon_{ijl}$ is the totally antisymmetric tensor, $P_B$ and $P_H$
are the non-helical and helical part of the spectrum for the ${\bf B}$ 
amplitude, respectively.
Scalar cosmological perturbations only couple to the non-helical part
of the spectrum and we shall therefore consider only $P_B$ in the following.

\subsection{Magnetic Energy Density}

As is clear from Eqs. (1-3), the EMT for PMF is quadratic in the field
amplitude. The PMF energy density spectrum is
\citep{KL}:
\begin{equation}
|\rho_B(k)|^2 =\frac{1}{128 \pi^2 a^8}\int d^3 p 
P_B(\vec p) P_B(|\vec k-\vec p|)
(1+\mu ^2)\,,
\label{edps}
\end{equation}
where
$\mu=\frac{\vec p (\vec k-\vec p)}
{p|\vec k-\vec p|}=\frac{k\cos \theta-p}{\sqrt(k^2+p^2-2kp\cos\theta)}$.
As for the two-point function in the coincidence limit,
for several physical spectral indexes and PMF configurations
such convolution is not finite. There are in general problems both on
large and short scales. Since the spectrum of the components of PMF EMT
are relevant for the final impact on cosmological perturbations and
CMB anisotropies, it is better to address this point in much more
detail with respect to what is present in literature.

\begin{figure}
\begin{tabular}{cc}
\includegraphics[scale=0.5]{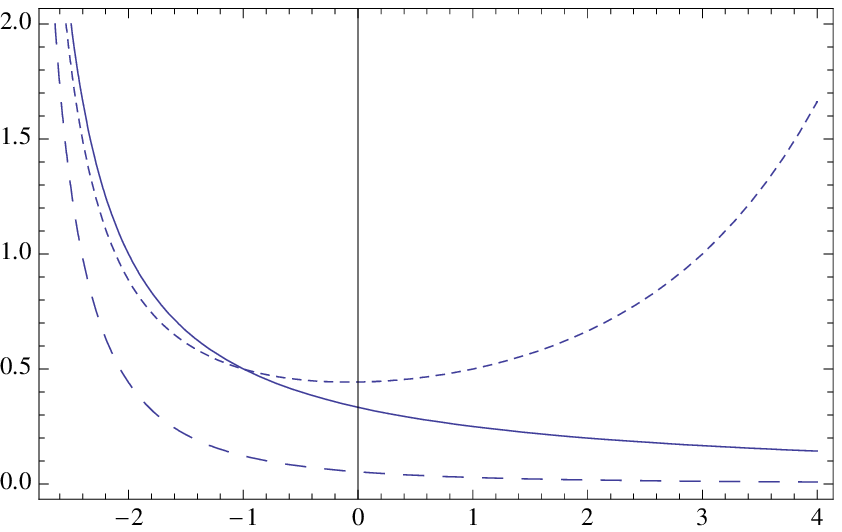}\includegraphics[scale=0.5]{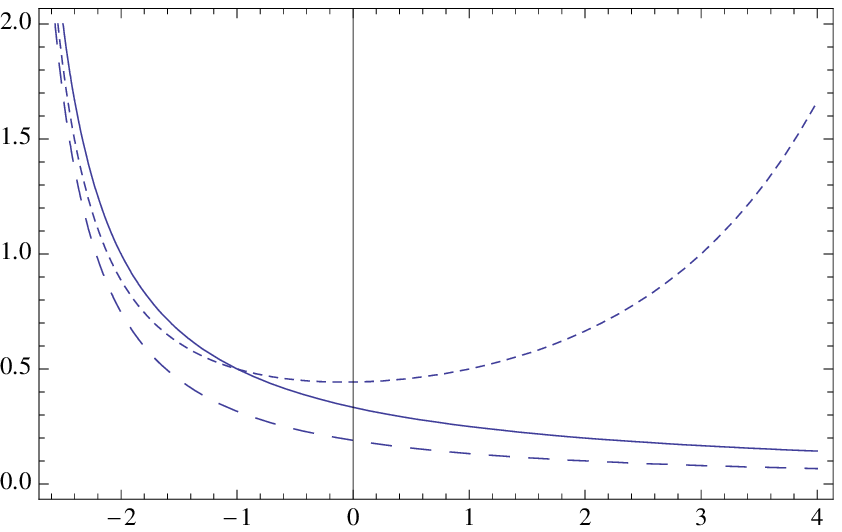}
\end{tabular}
\caption{Plots of the different ways of computing the magnetic power spectra 
(in units of $4 \pi A k_D^{n_B+3}/k_*^{n_B}$) versus $n_B$ for 
$k_S=k_D/2$ (left) and $k_S=k_D$ (right). 
$\langle B^2 \rangle$, $\langle B^2 \rangle_{k_S}$, 
$\langle B^2 \rangle_{{\rm cut} \,, k_S}$
are represented by solid, dotted and dashed lines, respectively.}
\label{comparisonrms}
\end{figure}

\begin{widetext}

The usual choice in the literature is to modify the scalar part
two-point function of Eq. (\ref{spectrum}) for zero helicity as \citep{KR}:
\begin{displaymath}
\langle\vec{B_i^*}(\vec{k})\vec{B_j}(\vec{k'})\rangle
= \left\{\begin{array}{ll}
\delta^3({\vec{k}}-{\vec{k'}})
(\delta_{ij}-\hat{k}_i\hat{k}_j) \frac{P_B(k)}{2}
& {\rm for}  \quad k < k_D \\ 0 & {\rm for}  \quad k > k_D
\label{spectrum2}
\end{array} \right. \,,
\end{displaymath}
\end{widetext}
with
\be
P_B (k) = A \left( \frac{k}{k_*} \right)^{n_B} \,,
\label{bps}
\ee
where $k_*$ is a reference scale.
With such choice the two-point function in the coincident limit 
(the mean square of the magnetic field) is:
\ba
\langle B^2 (x) \rangle &=& \int_{k<k_D} d^3 k
P_B (k) \nonumber \\ 
&=& \frac{4 \pi A}{n_B+3} \frac{k_D^{n_B+3}}{k_*^{n_B}} \,.
\label{Bsharp}
\ea 
It is also usual in the literature to
give the amplitude of $B$ at a given smearing scale $k_S$
by imposing a Gaussian filter :
\ba
\langle B^2 (x) \rangle_{k_S} &=& \int d^3 k 
P_B (k) e^{- k^2/k_S^2} \nonumber \\
&=& 2 \pi A
\frac{k_S^{n_B+3}}{k_*^{n_B}} \Gamma \left( \frac{n_B+3}{2} \right) \,.
\ea
By smearing the magnetic power spectrum and integrating for $k<k_D$, 
one gets: 
\ba
\langle B^2 (x) \rangle^{\rm cut}_{k_S} &=& \int_{k<k_D} d^3 k
P_B (k) e^{- k^2/k_S^2} \nonumber \\
&=& 2 \pi A
\frac{k_D^{n_B+3}}{k_*^{n_B}} \left[
\Gamma \left( \frac{n_B+3}{2} \right) \right. \nonumber \\
& & \left. - \Gamma \left( \frac{n_B+3}{2} \,, 
\frac{k_D^2}{k_S^2} \right) \right] \,,
\ea
where the incomplete Gamma function $\Gamma (... \,, ...)$ \cite{AS}
has been introduced.
Note how $n_B > -3$ in order to prevent infrared divergencies either in the
mean square field or the amplitude of the field smeared at a given scale. 
In the following by $\langle B^2 \rangle$ we mean the value given by 
Eq. (\ref{Bsharp}). Fig. (\ref{comparisonrms}) shows how 
$\langle B^2 (x) \rangle_{k_S}$ may be much larger than 
$\langle B^2 \rangle$ for $n_B > 0$.

\begin{figure}
\begin{tabular}{cc}
\includegraphics[scale=0.8]{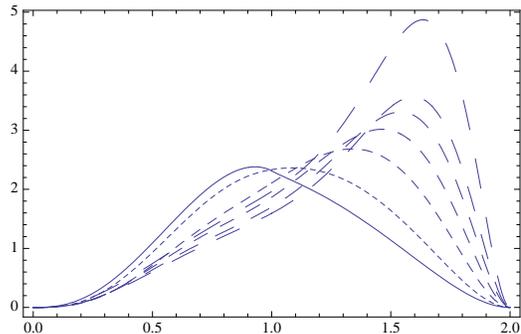}
\end{tabular}
\caption{Plot of magnetic energy density power spectrum 
$k^3 |\rho_B (k) |^2$ in units of $\langle B^2 \rangle^2/(1024 \pi^3)$ 
versus $k/k_D$ for different 
$n_B$ for fixed $\langle B^2 \rangle$. The different lines are for 
$n_B = -3/2, -1, 0, 1, 2, 3, 4$ ranging from the solid to the longest dashed.}
\label{energydensityps}
\end{figure}

The exact result for the Fourier convolution leading to the magnetic energy 
density Fourier square amplitude is one of the new main results of this paper. 
The convolution involves a double integral, one in the angle between 
$k$ and $p$ and one in the modulus of $p$. The integral in the angle, 
often omitted in the literature, is the reason for 
having the result for $|\rho_B(k)|^2$ non vanishing only for $k < 2 k_D$.
The detailed calculations for the energy density convolutions 
are given in Appendix A for several values of $n_B$. 
The generic behaviour for $k << k_D$ and $n_B > -3/2$ is white noise 
with amplitude 
\be
|\rho_B(k)|^2 \simeq \frac{A^2k_D^{2n+3}}{16 \pi k_*^{2n} (3+2n_B)} 
\label{infrareden}
\ee
and then goes to zero for $k=2k_D$, which is a result obtained by 
performing correctly the integral.
The pole for $n_B = -3/2$ 
in Eq. (\ref{infrareden}) is replaced by a logarithmic diveregence in $k$ 
in the exact result; for $n_B < -3/2$ the spectrum is no more white noise for 
$k << k_D$. Fig. (\ref{energydensityps}) shows the dependence of 
$k^3 |\rho_B(k)|^2$
on $n_B$ at fixed $\langle B^2 \rangle$. 

\begin{figure}
\begin{tabular}{cc}
\includegraphics[scale=0.8]{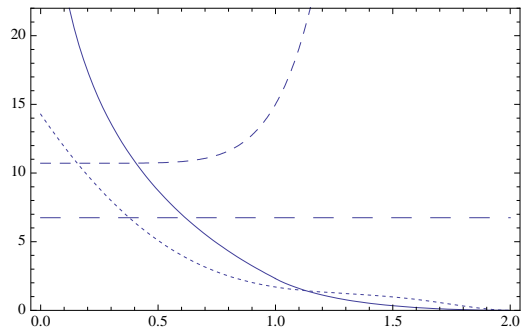}
\end{tabular}
\caption{Comparison of magnetic energy density convolution
$k_D^3 |\rho_B (k) |^2$ obtained in this paper (dotted, solid) 
in units of $\langle B^2 \rangle^2/(1024 \pi^3)$ and the one 
in Eq. (\ref{KR_convolution}) 
(dashed, long-dashed) versus $k/k_D$ for $n_B=2,-3/2$ with fixed 
$\langle B^2 \rangle$.}
\label{comparison_convolution}
\end{figure}

Our result 
are different from the one reported 
in the literature \cite{KR}, which is 
\be
|\rho_B(k)|^2_{KR} = \frac{3 A^2k_D^{2n_B+3}}{64 \pi k_*^{2n_B} (3+2n_B)} 
\left[ 1 + \frac{n_B}{n_B + 3} \left( \frac{k}{k_D} \right)^{2n_B + 3} 
\right] \,,
\label{KR_convolution}
\ee
and is not limited in $k$. In Fig. (\ref{comparison_convolution}) 
we show the difference between the literature result \cite{KR} 
and our result for $n_B=2,-3/2$.

\begin{widetext}
\subsection{Lorentz Force}

As is clear from previous sections, we also need the Lorentz force
\ba
|L (k)|^2 =\frac{1}{128 \pi^2 a^8}\int &d^3 p& P_B(p) \, P_B(|{\mathbf{k}}
-{\mathbf{p}}|) [1 + \mu^2 + 4 \gamma \beta(\gamma \beta - \mu)] \,,
\label{spectrum_LF}
\ea
and the magnetic anisotropic stress 
\be
|\sigma_B (k)|^2 =\frac{1}{288 \pi^2 a^8}\int d^3 p \, P_B(p) \, 
P_B(|{\mathbf{k}}
-{\mathbf{p}}|) [9(1-\gamma^2)(1-\beta^2)-6(1+\gamma\mu\beta-\gamma^2
-\beta^2)(1+\mu^2)] \,,
\label{spectrum_AS}
\ee
where $\gamma= \hat k \cdot \hat p$,
$\beta= \vec k \cdot (\vec k -\vec p)/(k |\vec k-\vec p|)$ and $\mu =
\vec p \cdot (\vec k - \vec p)/(p|\vec k -\vec p|)$. 
\end{widetext}

\begin{figure}
\begin{tabular}{cc}
\includegraphics[scale=0.8]{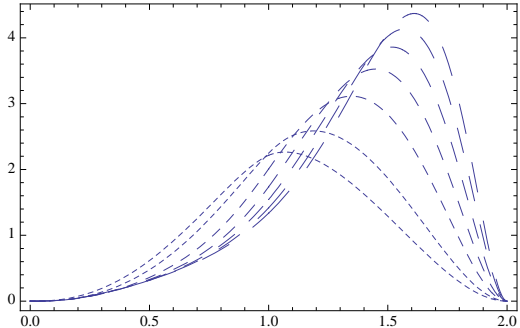}
\end{tabular}
\caption{Plot of the Lorentz force power spectrum 
$k^3 |L (k) |^2$ in units of $\langle B^2 \rangle^2/(1024 \pi^3)$ 
versus $k/k_D$ for different 
$n_B$ for fixed $\langle B^2 \rangle$. The different lines are for     
$n_B = -3/2, -1, 0, 1, 2, 3$ ranging from the solid to the longest dashed.}
\label{lorentzps}
\end{figure}

We decide to compute the spectrum of the Lorentz force and obtain the 
anisotropic stress by Eq. (\ref{stress_B}). The exact computation for the 
Lorentz force power spectrum is given in Appendix B for several values 
of $n_B$. A term $- \rho_B$ can be easily identified in Eq. (\ref{LFS}); 
since we know from the exact computation that 
the integral of $P_B(p) \, P_B(|{\mathbf{k}}-{\mathbf{p}}|) (1+\mu^2)$
is larger than the remaining piece in Eq. (\ref{spectrum_LF}) we chose the 
the signs for $\rho_B (k)$ and $L (k)$ as opposite.

\begin{figure}
\begin{tabular}{cc}
\includegraphics[scale=0.8]{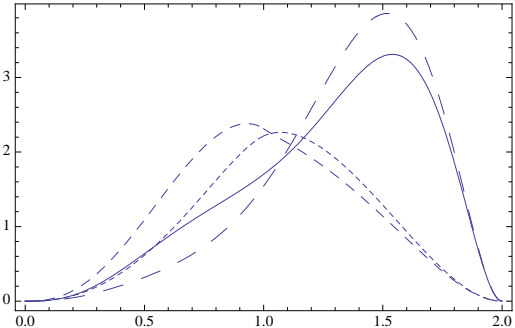}
\end{tabular}
\caption{Comparison of the magnetic energy density and 
Lorentz force power spectra
versus $k/k_D$ for fixed $\langle B^2 \rangle$. The solid 
(medium dashed) and long-dashed (short-dashed)
lines are respectively for $k^3 |\rho_B (k) |^2$ and $k^3 |L(k) |^2$
for $n_B = 2$ ($n_B=-3/2$).}
\label{comparison_rhovsL}
\end{figure}

Fig. (\ref{lorentzps}) shows the dependence of 
$k^3 |L (k)|^2$ on $n_B$ at fixed $\langle B^2 \rangle$. 
Fig. (\ref{comparison_rhovsL}) compares the approximation 
$L(k) \simeq - \rho_B(k)$ suggested in 
Ref. \cite{KR} with the exact calculation. As can be checked in Appendix B, 
our exact calculations for the values of $n_B$ studied here show that 
\be
|L(k)|^2 \simeq \frac{11}{15} |\rho_B (k)|^2 \quad {\rm for} \quad k << k_D \,.
\label{infrared_relation}
\ee

\section{Results for Cosmological Perturbations}

In order to study the effects of a SB of PMFs on CMB anisotropies 
and matter power spectrum we modified the CAMB Einstein-Boltzmann code 
\cite{CAMB} (June 2006 version) 
by introducing the PMF contribution in the Einstein equations, 
in the evolution equation for baryons and initial conditions, 
along Eqs. (\ref{Einsteineqs},\ref{baryonseq},\ref{initialconds}). 

\begin{figure}
\begin{tabular}{cc}
\includegraphics[scale=0.25]{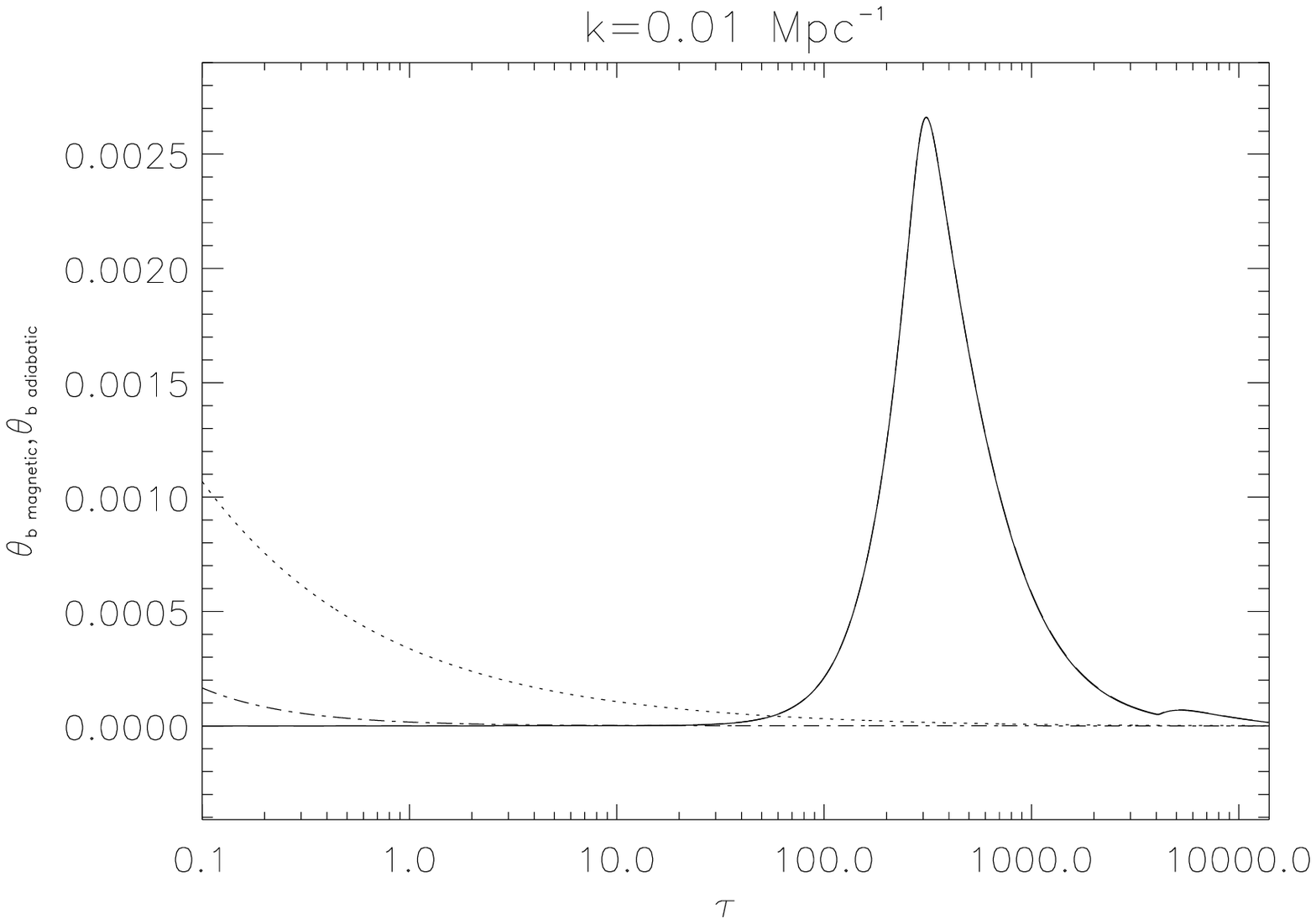}\includegraphics[scale=0.25]{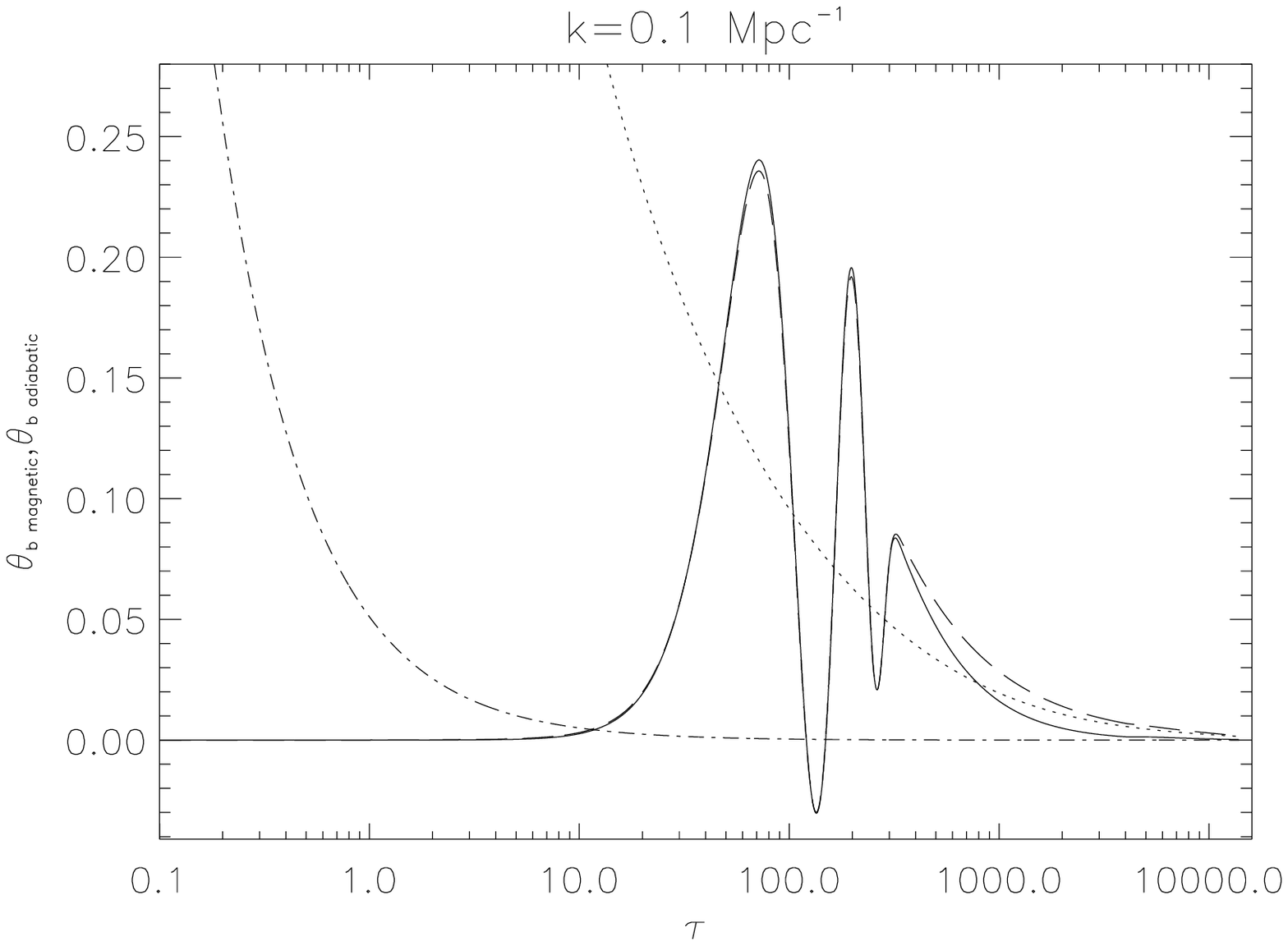} \\ 
\includegraphics[scale=0.25]{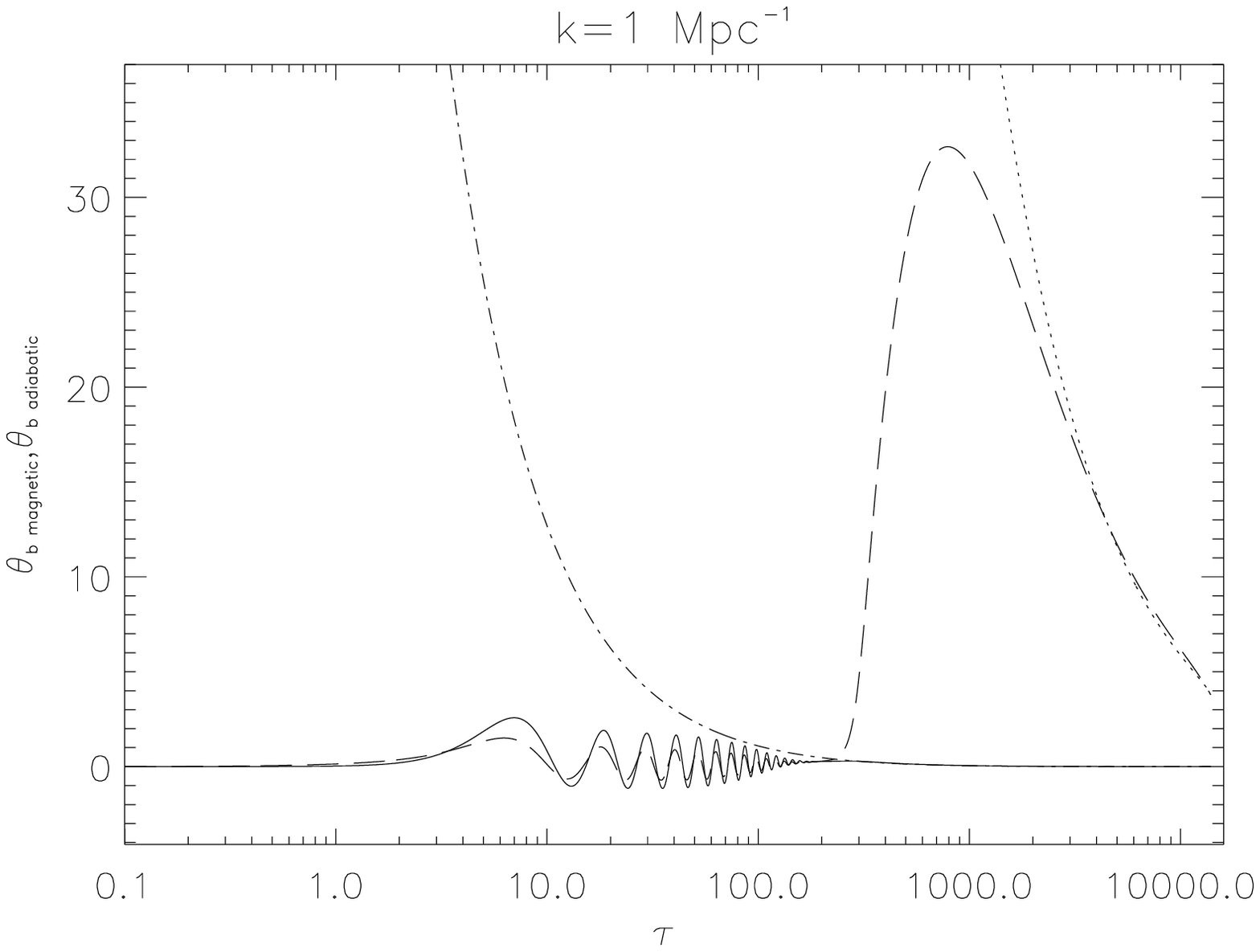}\includegraphics[scale=0.25]{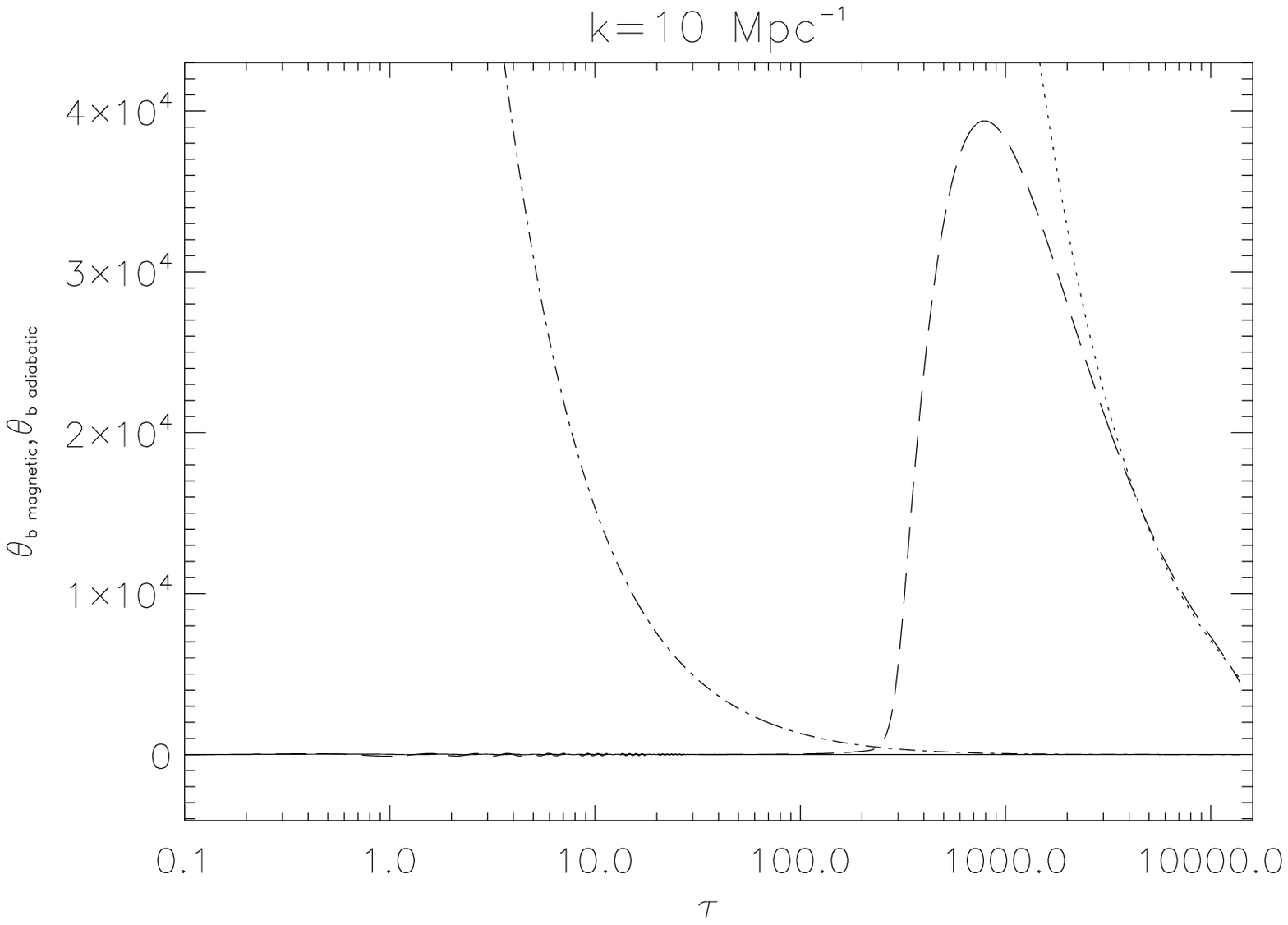}
\end{tabular}
\caption{Evolution of baryons velocity for $4$ different wavenumbers 
with (dashed) and without (solid) PMF. $k^2 L /\rho_b$ (dot-dashed line) and 
the solution $\theta_b^{\rm late}$ (dotted line) are also plotted: 
note how the numerics agree with $\theta_b^{\rm late}$ at late times.
The cosmological parameters of the flat $\Lambda CDM$ model are 
$\Omega_b h^2 = 0.022$, 
$\Omega_c h^2 = 0.123$, $\tau = 0.04$, $n_s=1$, 
$H_0 = 72 \, {\rm km \, s}^{-1} \, {\rm Mpc}^{-1}$. }
\label{fig_thetab}
\end{figure}

\begin{figure}
\begin{tabular}{cc}
\includegraphics[scale=0.25]{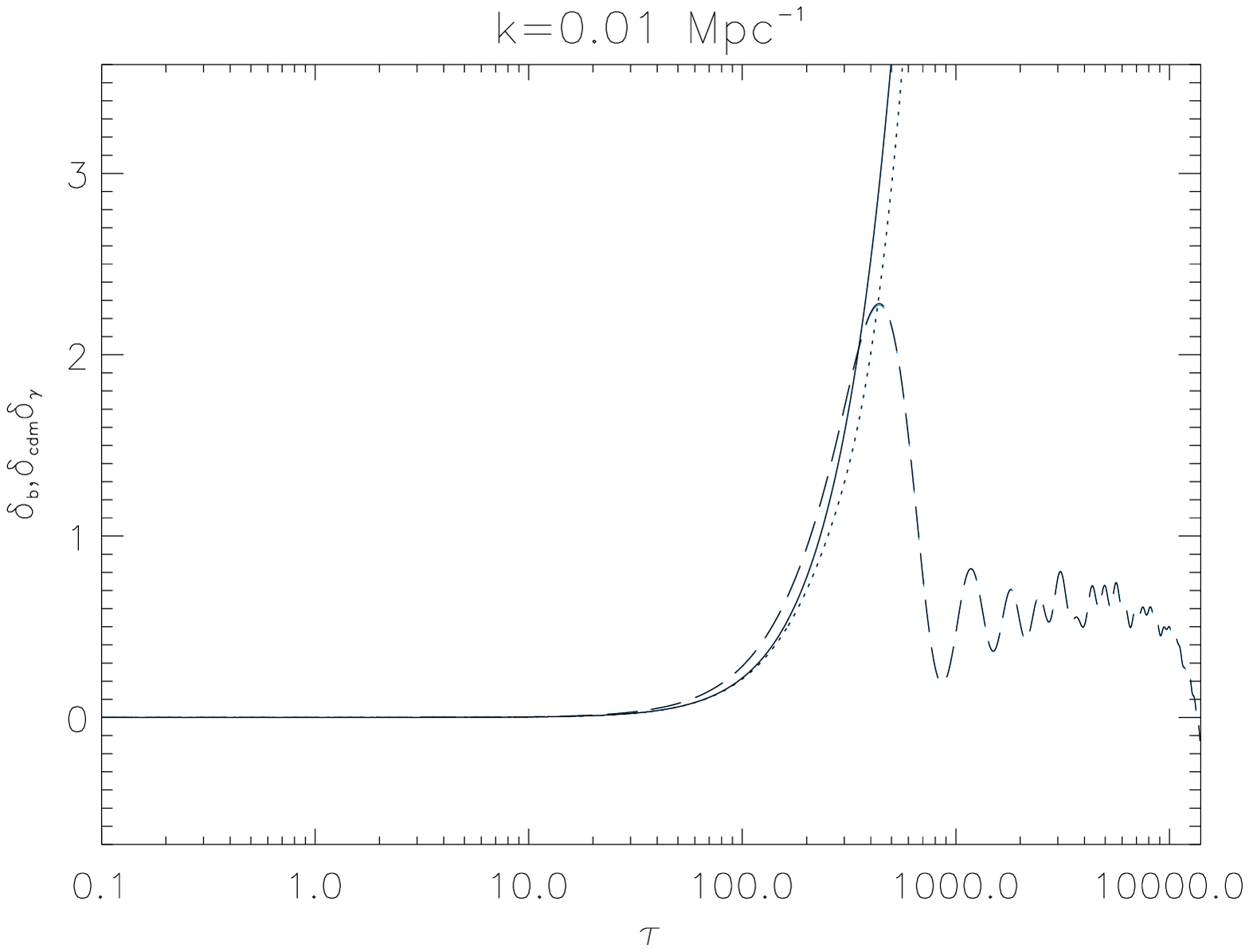}\includegraphics[scale=0.25]{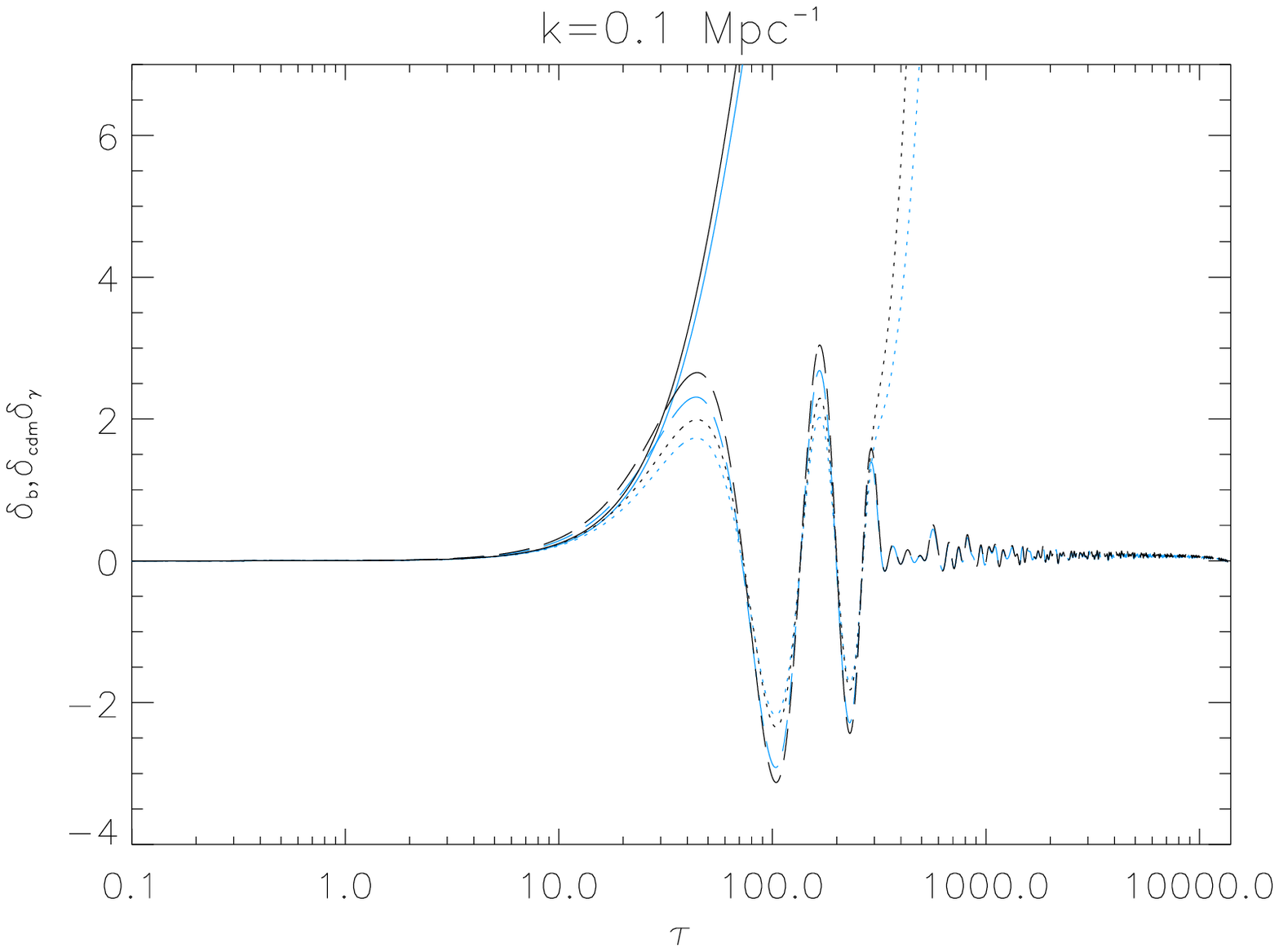}\\
\includegraphics[scale=0.25]{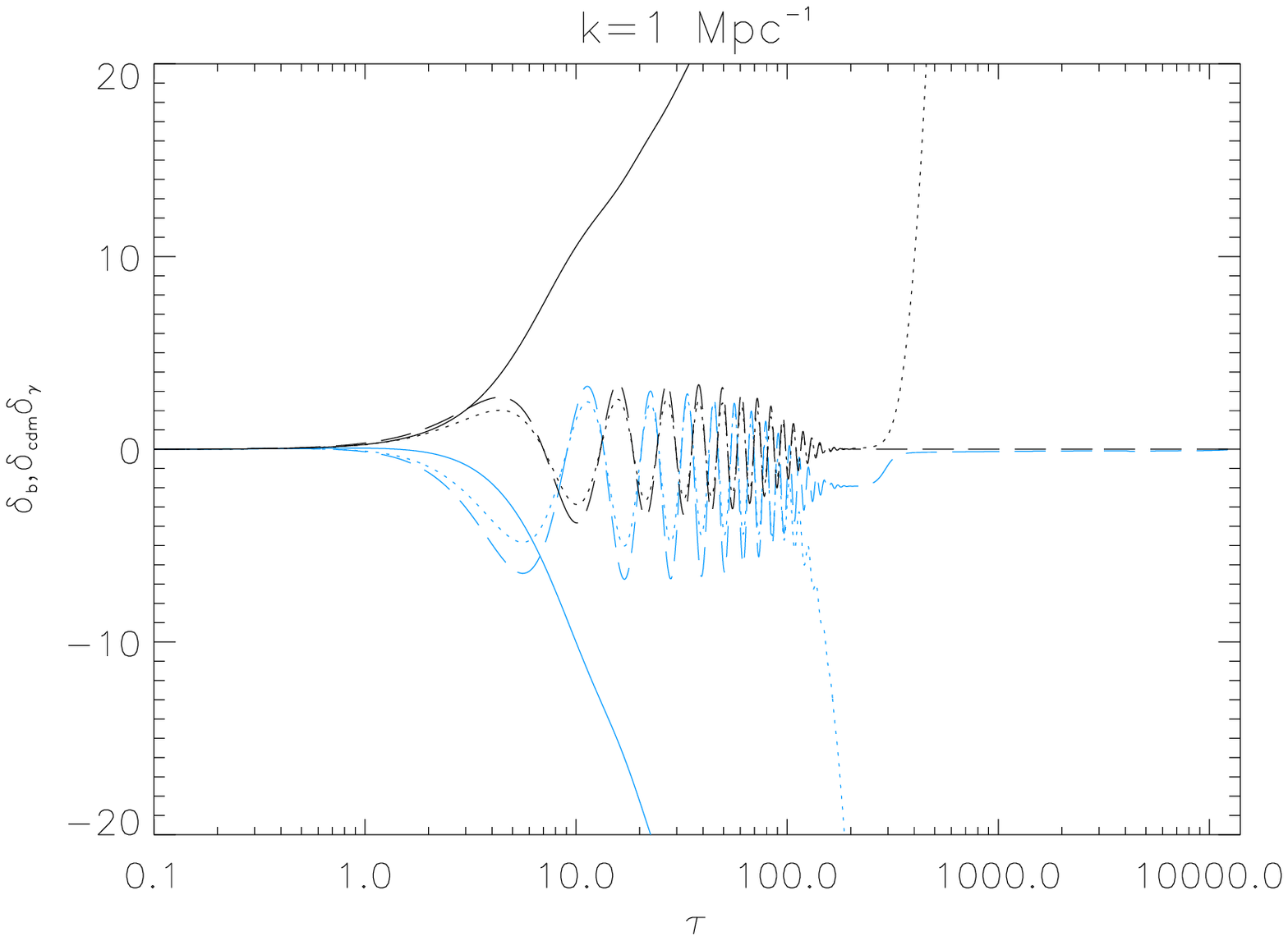}\includegraphics[scale=0.25]{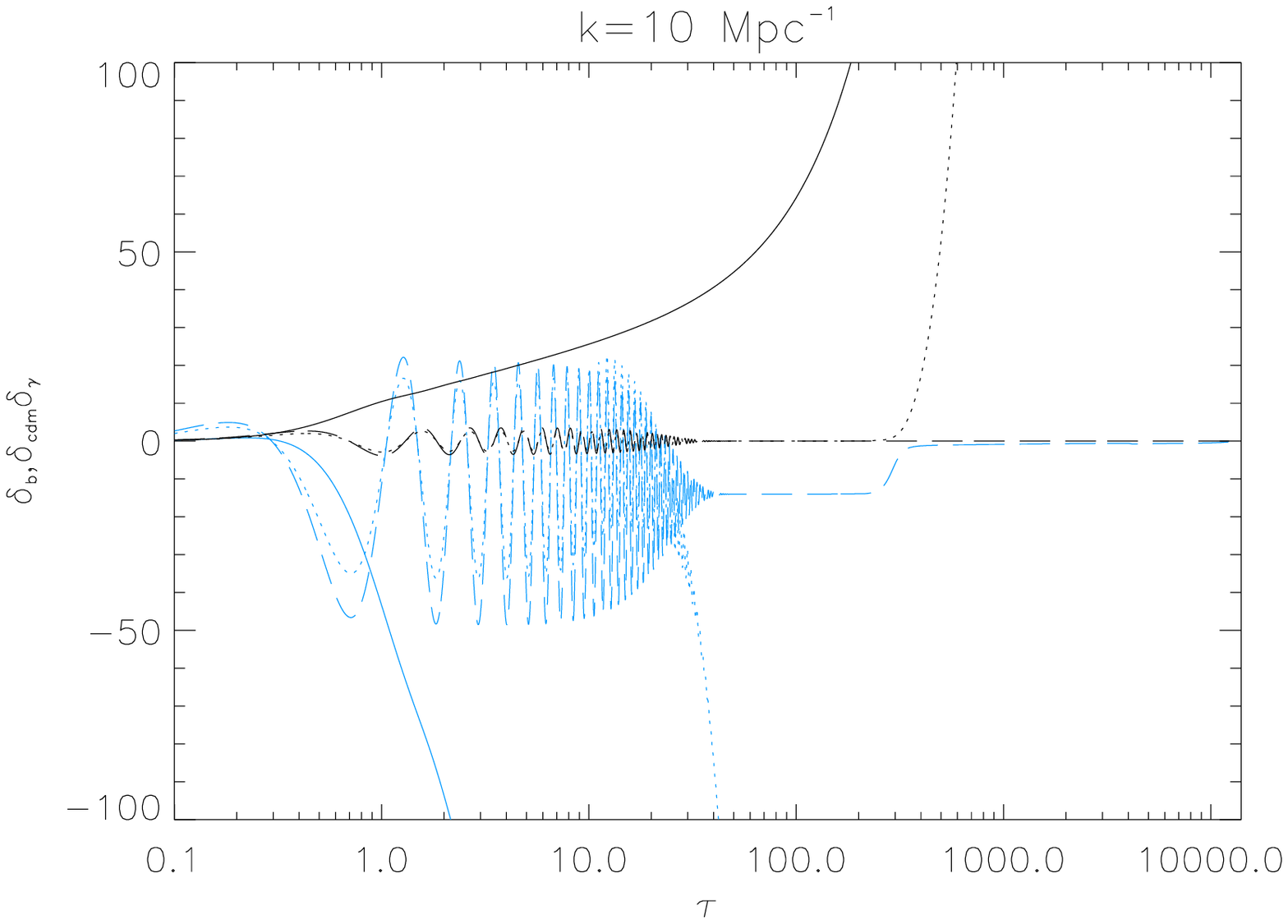}
\end{tabular}
\caption{Evolution of baryons (dotted), CDM (solid) and 
photons (dashed) density contrast for $4$ different wavenumbers
with fully correlated (blue) and without (black) PMF. 
The cosmological parameters are the same of Fig. (\ref{fig_thetab}).
}
\label{fig_deltab}
\end{figure}

\begin{figure}
\begin{tabular}{cc}
\includegraphics[scale=0.25]{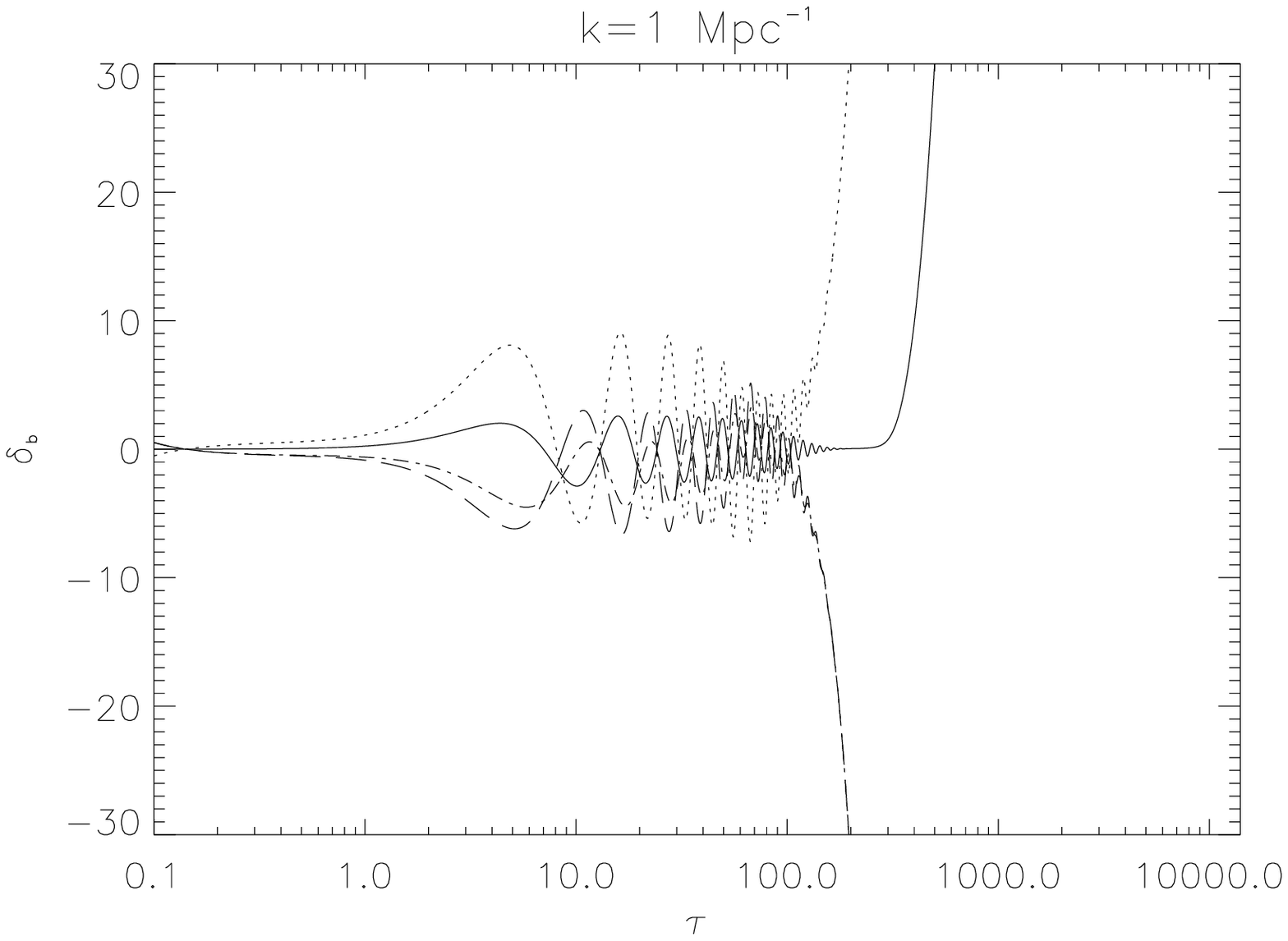}\includegraphics[scale=0.25]{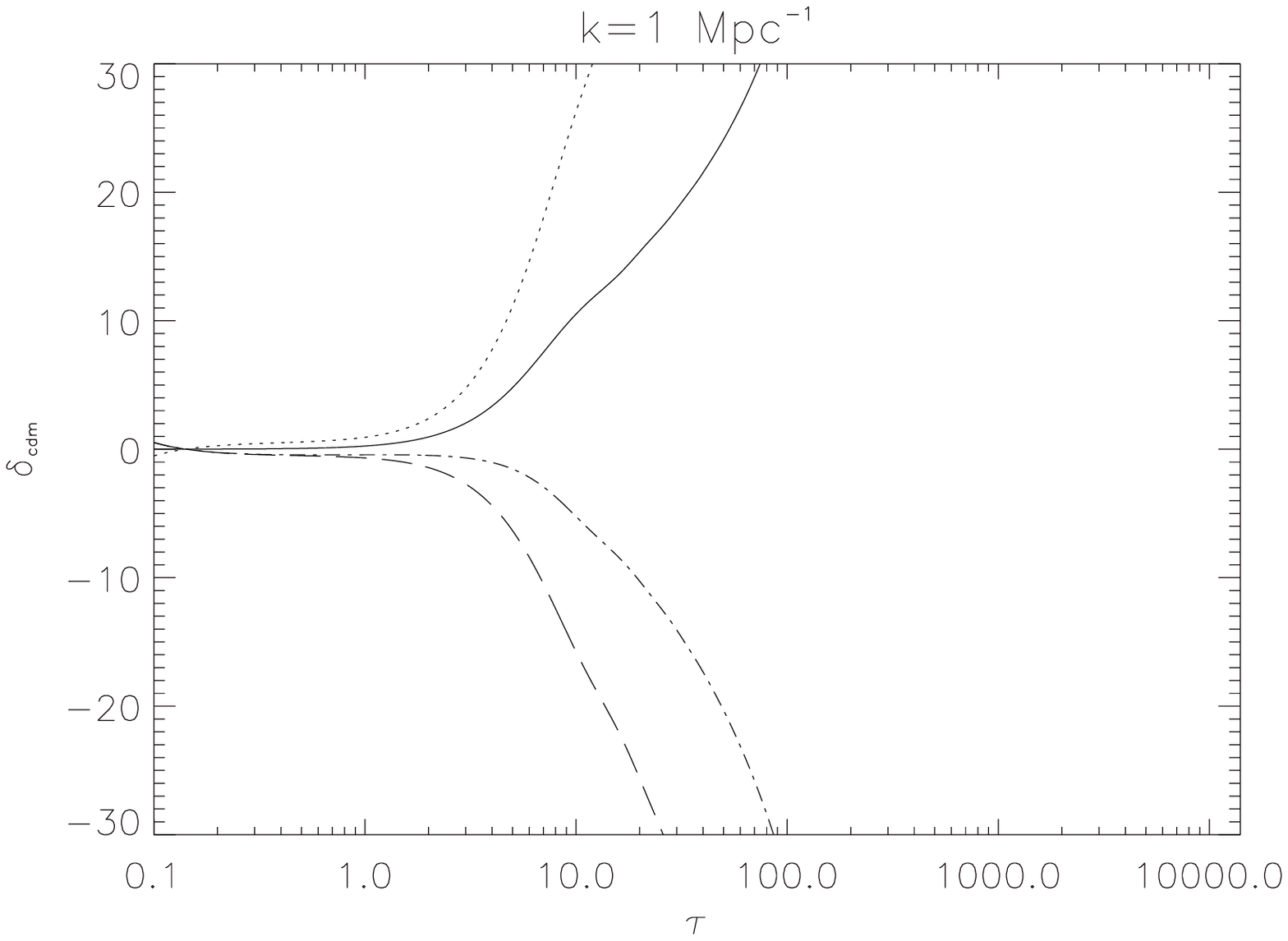}
\end{tabular}
\caption{Time evolution of baryons (left) and CDM (right) density contrasts
with vanishing PMF (solid), fully correlated (dashed), fully anti-correlated 
(dotted) and purely magnetic initial conditions
(dot-dashed). The other cosmological parameters are the same as Fig. 
(\ref{fig_thetab}).}
\label{fig_matter}
\end{figure}

\begin{figure}
\includegraphics[scale=0.25]{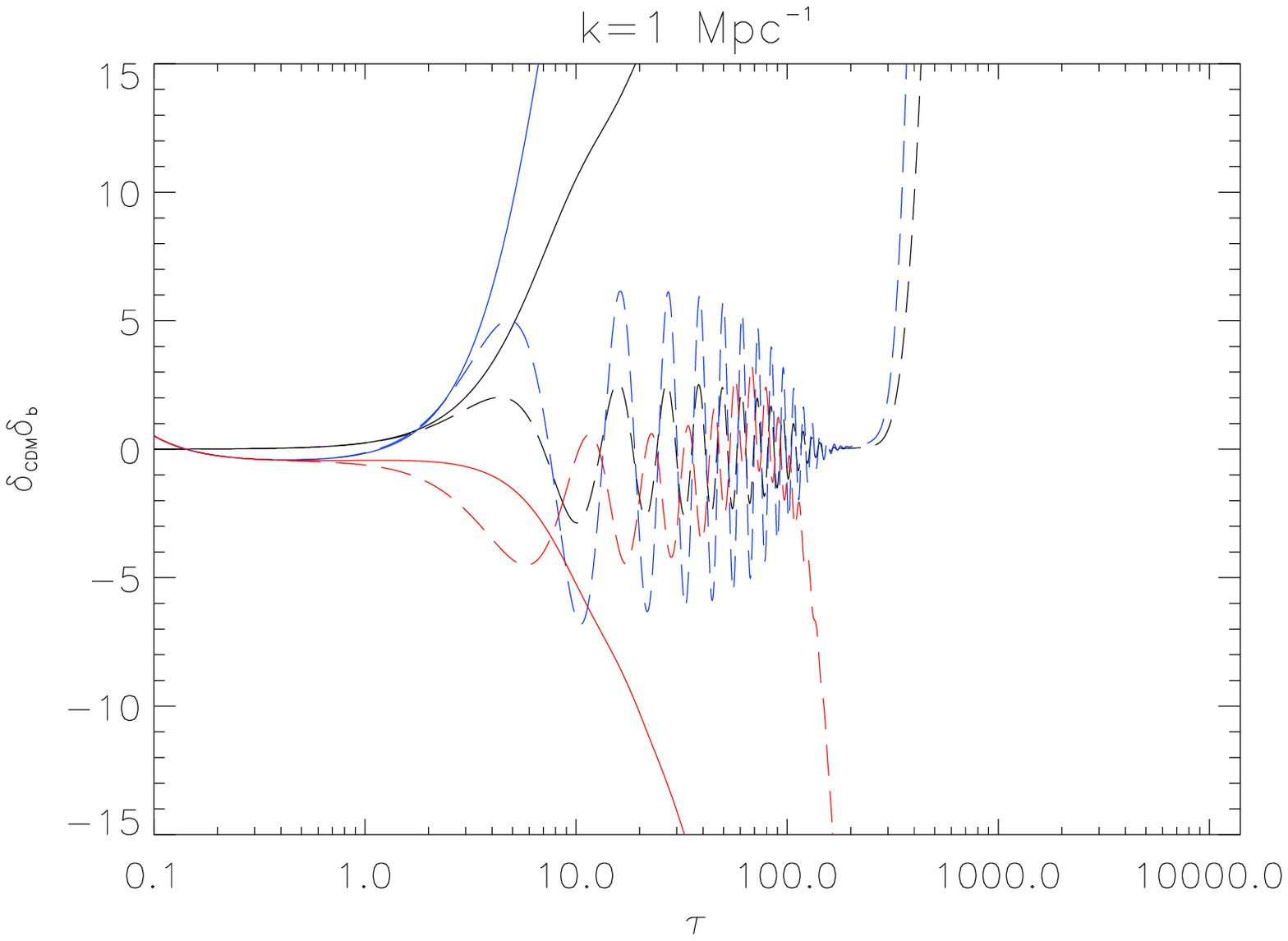}\includegraphics[scale=0.25]{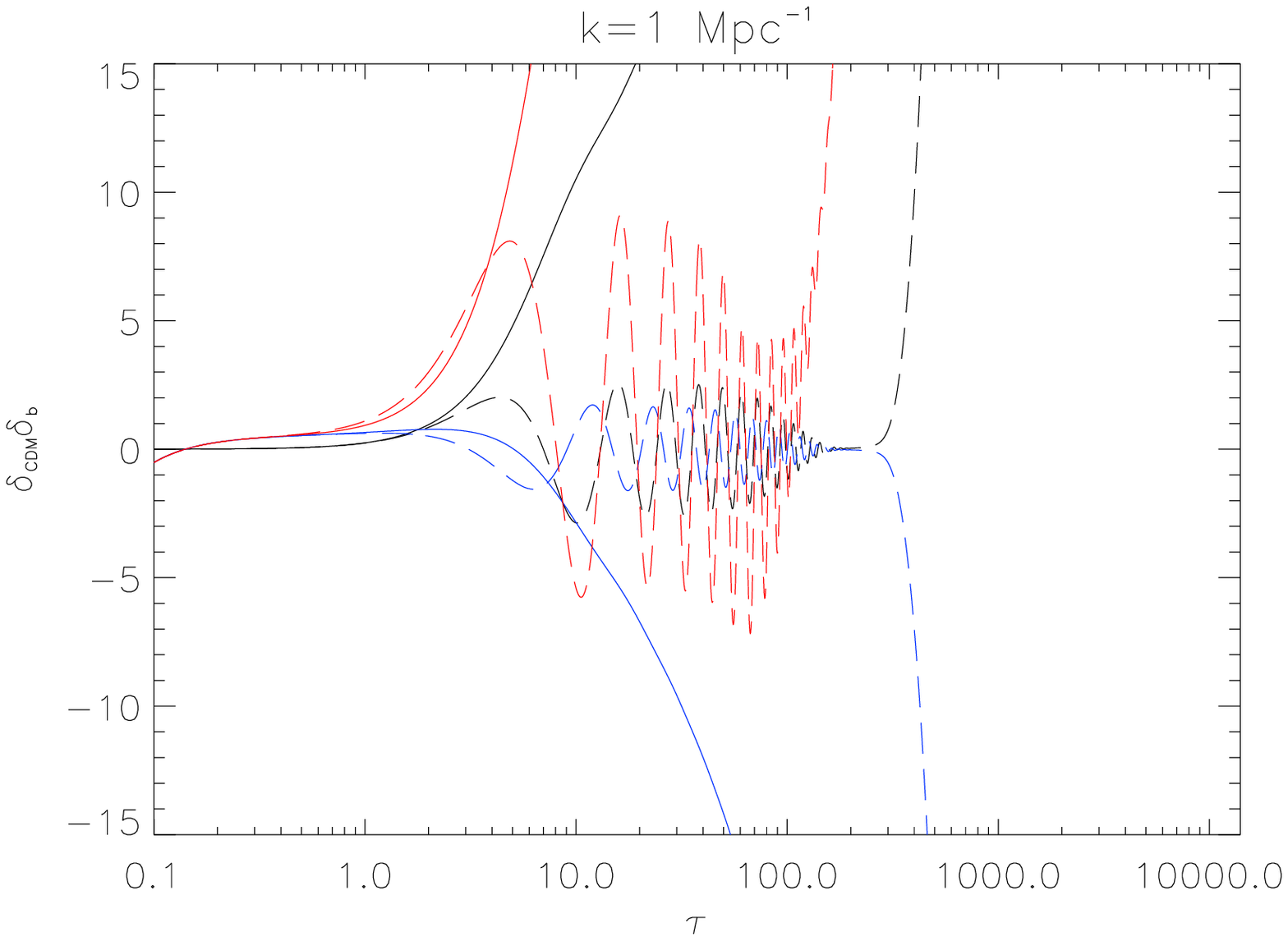}
\caption{Time evolution of baryons (dashed) and CDM (solid) density contrasts for purely adiabatic with vanishing PMF (black), 
fully correlated (left panel) and anti-correlated (right panel) PMF with vanishing (blue) and non vanishing 
(red) Lorentz force for $k=1 \, {\rm Mpc}^{-1}$. 
These figures show clearly that the Lorentz 
force and the gravitational contribution are of opposite sign, and the Lorentz term is more important.
The cosmological parameters are the same of Fig. (\ref{fig_thetab}).}
\label{fig_lorentzvsgravity}
\end{figure}

We note that implementing the baryons evolution as from 
Eq. (\ref{baryonseq}), the MHD approximation in a globally neutral plasma 
is used up to the present time: this makes the Lorentz term non-vanishing 
up to the present time. Although the term $L/\rho_b$ 
in Eq. (\ref{baryonseq}) decreases with time, its effect on the baryon 
velocity is crucial. Much later than the larger between the 
decoupling time and the time at which sound speed of baryons is effectively 
zero, baryons velocity can be approximated as:
\be
\theta_b^{\rm late} \simeq - k^2 \left( \frac{L a}{\rho_b} \right)
\frac{\tau}{a}
\label{baryonseq}
\ee
during the matter dominated era. Our modified Einstein-Boltzmann code 
reproduces correctly this asymptotic regime for different wavelengths, 
as can be seen by Fig. (\ref{fig_thetab}). The corresponding effects 
on the density contrasts for the same wavelengths are shown 
in Fig. (\ref{fig_deltab}). 
In Fig. (\ref{fig_matter}) the effects due to 
the pure magnetic mode and due to the correlation with the adiabatic 
mode are shown. Fig. (\ref{fig_lorentzvsgravity}) displays the importance 
of the Lorentz term compared to the purely gravitational effect.

\section{Results for CMB Temperature and Polarization Power Spectra}

\begin{figure}
\includegraphics[scale=0.5]{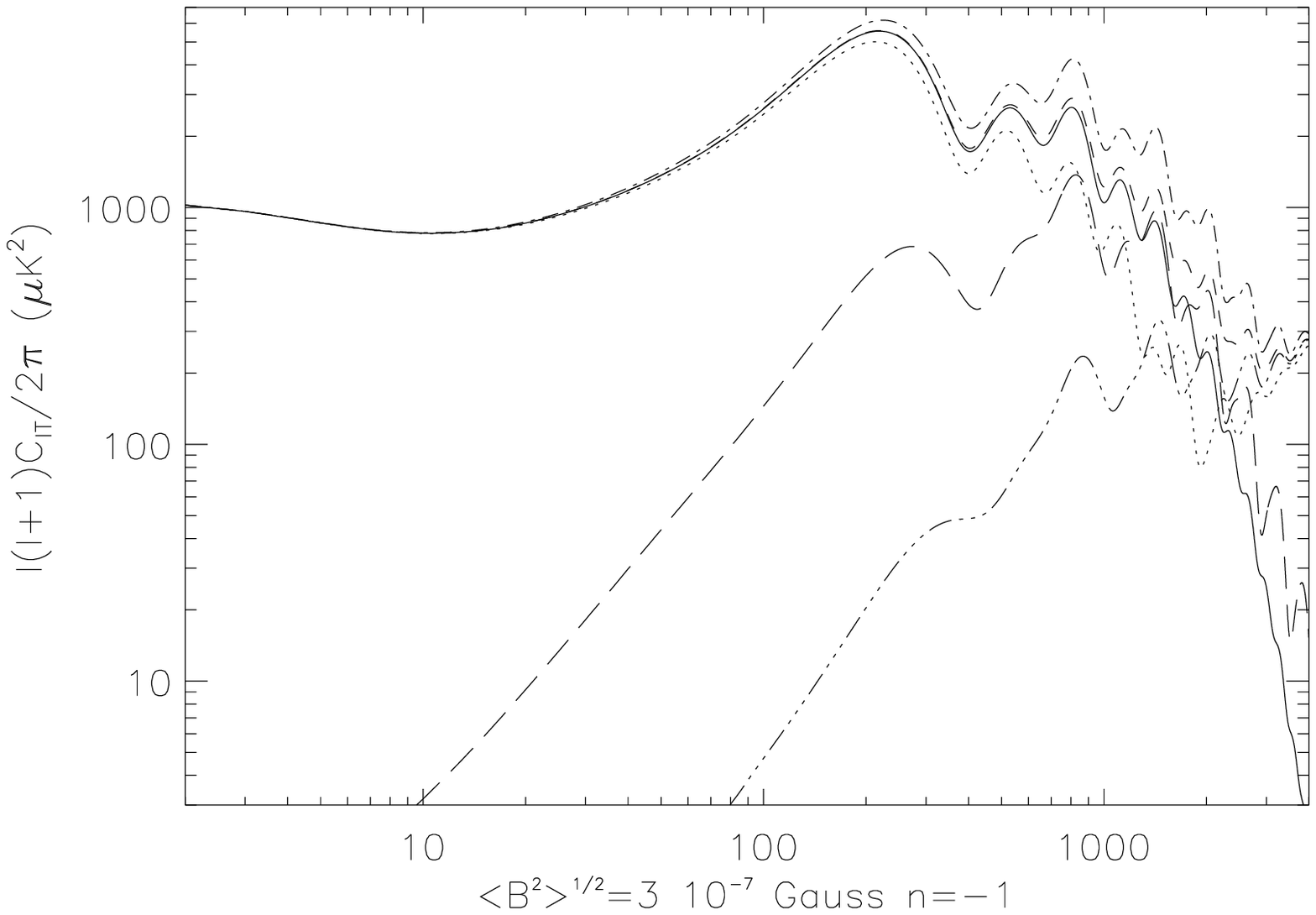}
\includegraphics[scale=0.5]{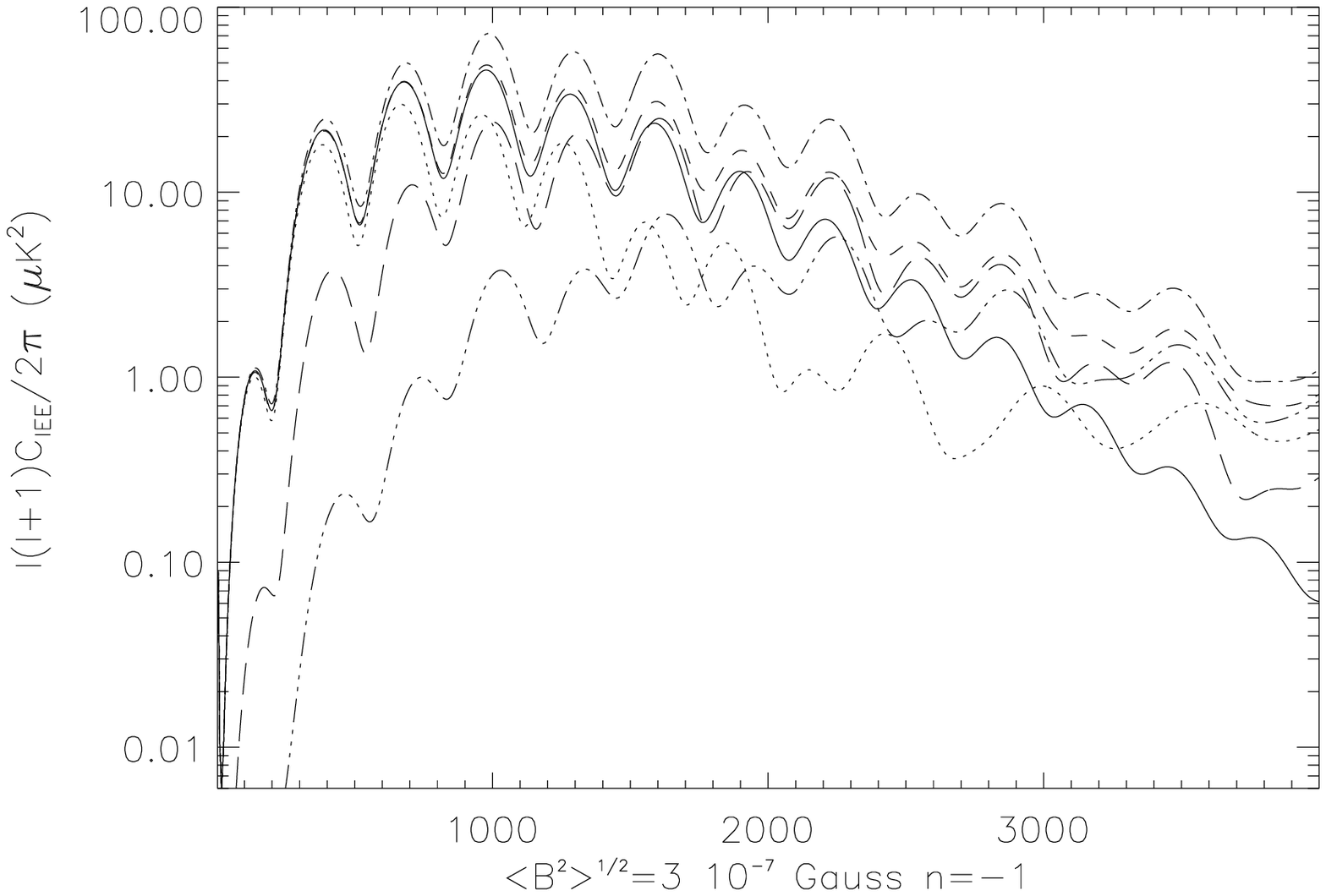}
\includegraphics[scale=0.5]{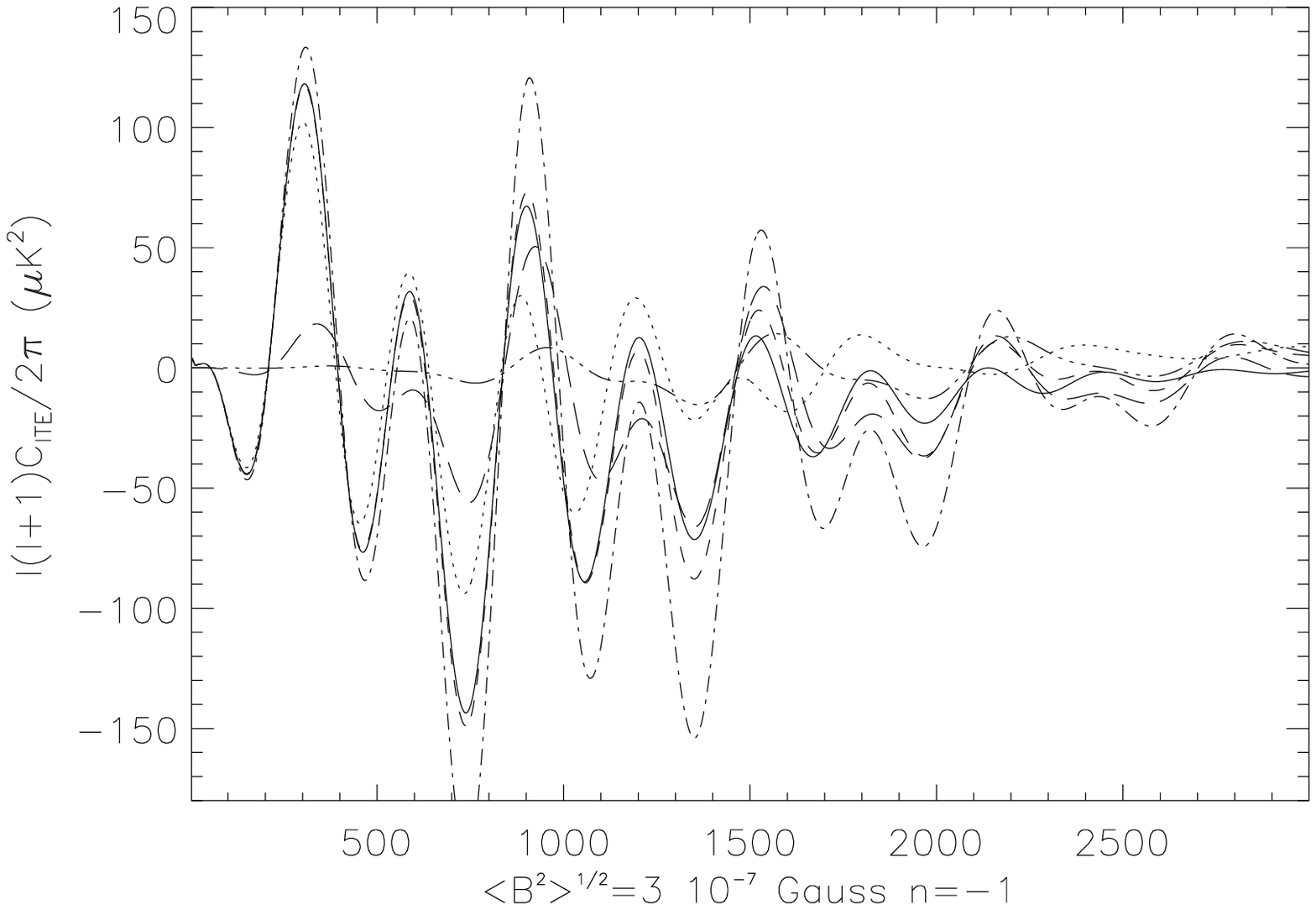}
\caption{CMB temperature angular power spectra obtained with 
$\sqrt{\langle B^2 \rangle} = 3 \times 10^{-7} \, {\rm Gauss}$, $n_B = -1$, 
$k_D = \pi$ in comparison with the adiabatic spectrum with vanishing PMF 
(solid line): TT, EE, TE are displayed in the top, middle, bottom panel, 
respectively. 
The purely magnetic, correlation, fully correlated, 
fully anti-correlated and uncorrelated spectra 
are represented as triple dotted - dashed, 
dashed, dotted, dot - dashed and long dashed lines, respectively. 
The other cosmological parameters 
are the same as Fig. (\ref{fig_thetab}).}
\label{various_contributions}
\end{figure}

\begin{figure}
\includegraphics[scale=0.5]{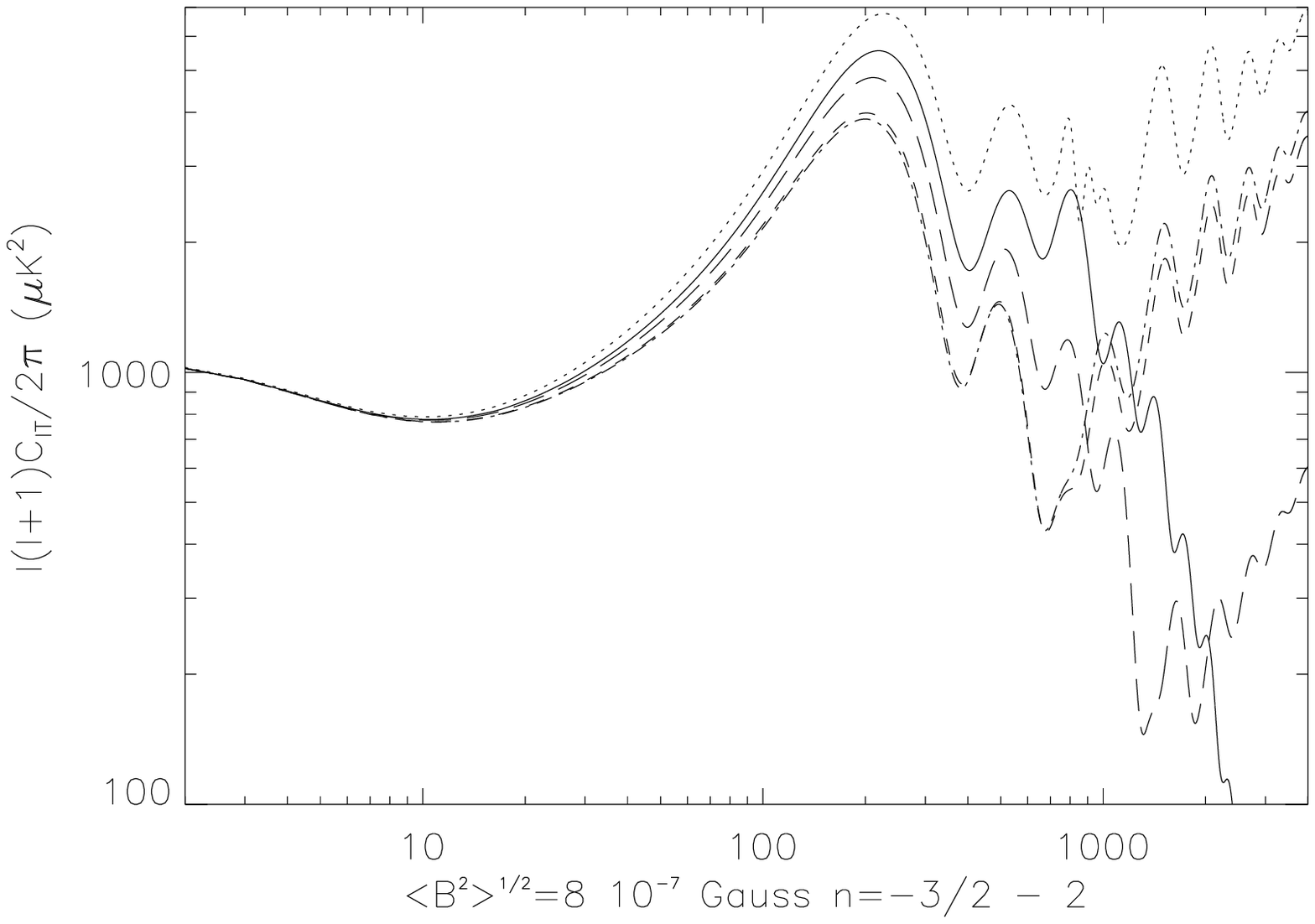}
\includegraphics[scale=0.5]{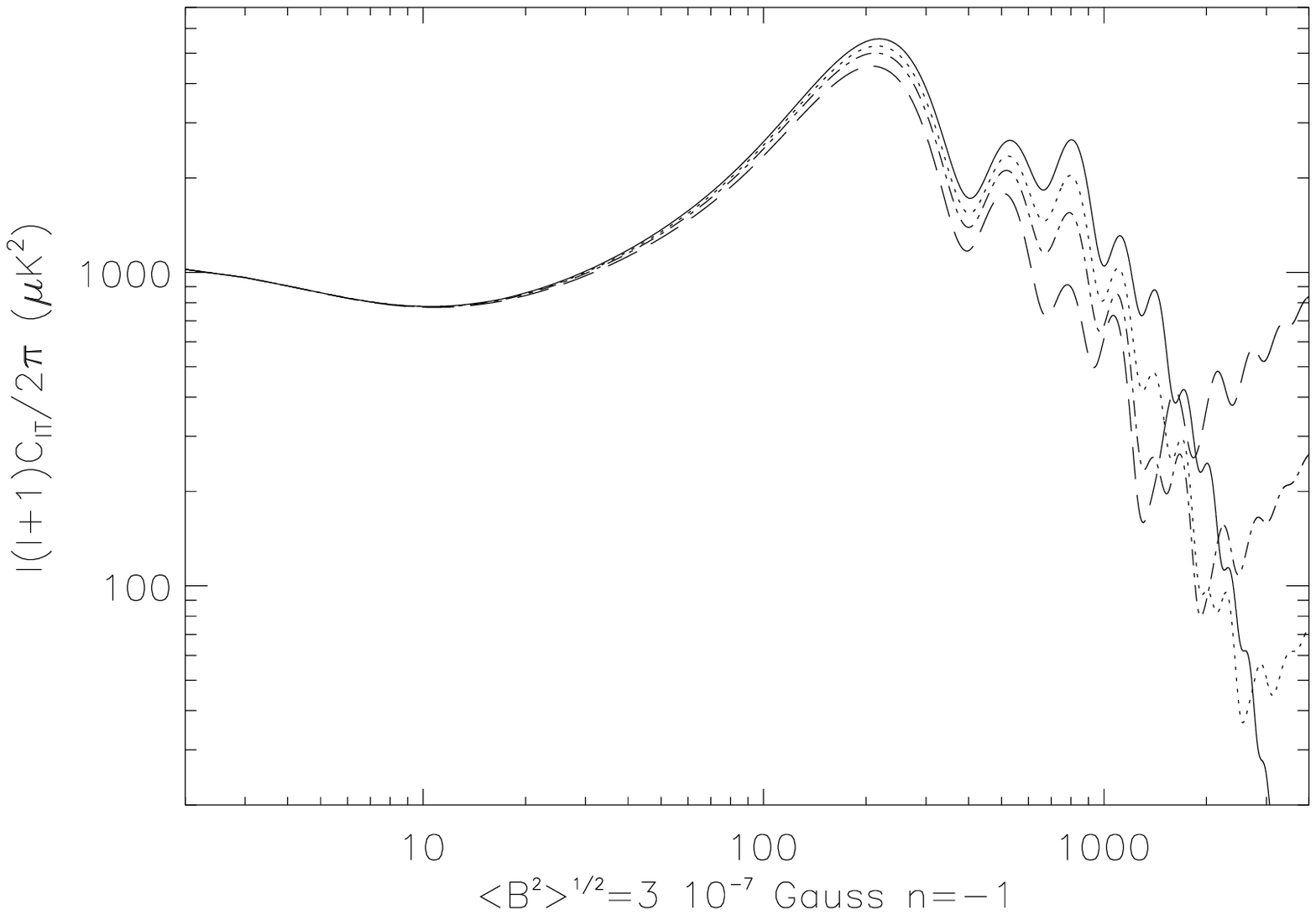}
\caption{In the top panel 
variation of the CMB temperature angular power spectrum 
with $n_B$ 
in comparison to the case with vanishing PMF (solid line). In the top figure 
$\sqrt{\langle B^2 \rangle} = 8 \times 10^{-7} \,
{\rm Gauss}$, $k_D = 2 \pi$ and fully correlated initial conditions 
are considered. The spectral indexes plotted are 
$n_B = -3/2 \,, -1 \,, 1 \,, 2$ (dotted, dot-dashed, dashed, long-dashed 
lines, respectively). In the bottom panel, variation of the CMB angular 
power spectrum with $k_D$ in comparison to the case with vanishing PMF 
(solid line). In the bottom figure
$\sqrt{\langle B^2 \rangle} = 3 \times 10^{-7} \,
{\rm Gauss}$, $n_B = -1$ and 
$k_D = 2 \pi \,, \pi \,, \pi/2$ (dotted, dot-dashed, dashed, respectively).
In both panels the initial conditions are fully correlated
and the other cosmological parameters are the same as 
Fig. (\ref{fig_thetab}).}
\label{varying_nB}
\end{figure}

In this section we show the results on the CMB temperature and polarization 
pattern obtained by our modifications of the CAMB code. 
Fig. (\ref{various_contributions}) shows 
the various contributions to the total CMB temperature and polarization 
angular power spectra from the 
pure magnetic mode and its correlation with the adiabatic mode. 
Fig. (\ref{varying_nB}) shows the dependence of the total temperature power 
spectrum on the spectral index $n_B$.

As is clear from the previous section, the Lorentz force of 
a fully correlated magnetic 
contribution decreases the density contrasts and therefore the CMB APS
in an intermediate range of multipoles before high $\ell$ increase.

\begin{figure}
\includegraphics[scale=0.25]{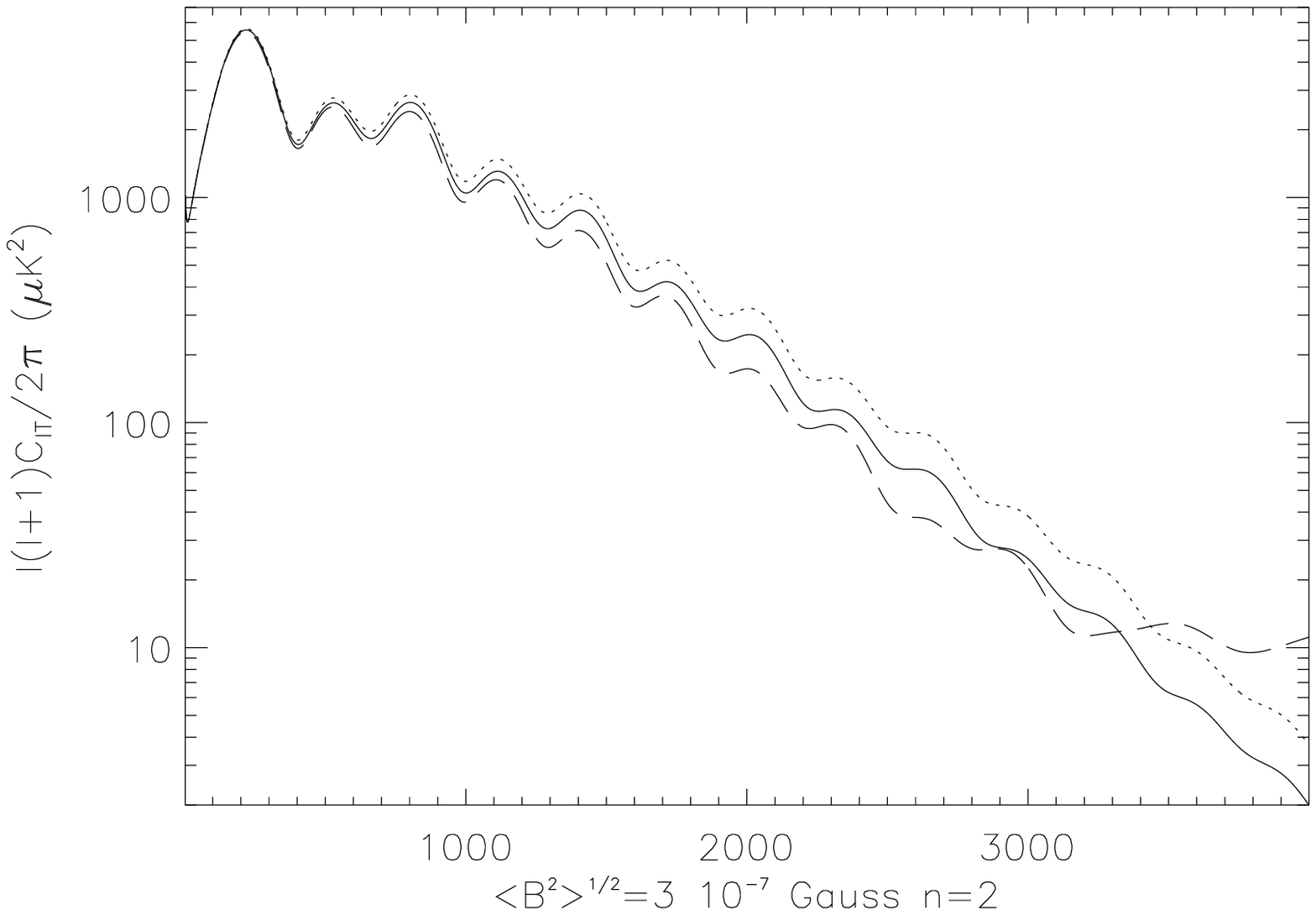}\includegraphics[scale=0.25]{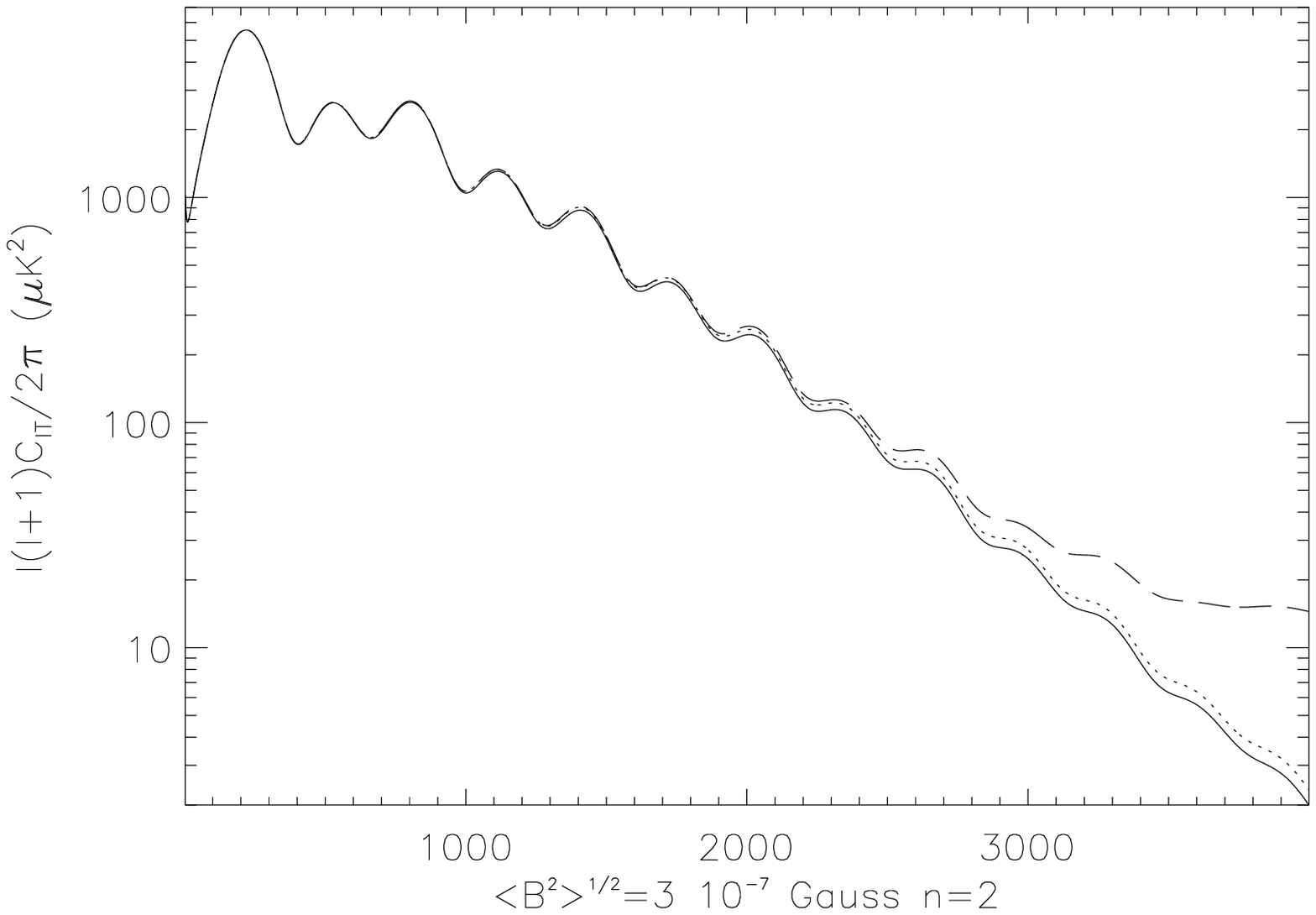}
\caption{In the left panel CMB temperature power spectrum obtained with fully correlated
PMF with (dashed line) and without (dotted line) Lorentz term in comparison
with the vanishing PMF (solid line). As is clear from the previous section, 
the Lorentz force of a fully correlated magnetic
contribution decreases the density contrasts and therefore there is 
range in which the CMB TT APS is decreased respect to the adiabatic case.
In the right panel the same figure with uncorrelated spectra.
In the figures
$\sqrt{\langle B^2 \rangle} = 3 \times 10^{-7} \,
{\rm Gauss}$, $k_D = 2 \pi$ and $n_B=2$ are considered.
The other cosmological parameters are the same as Fig. (\ref{fig_thetab}).}
\label{CMB_lorentz}
\end{figure}

\section{Results for the Matter Power Spectra}

\begin{figure}
\includegraphics[scale=0.5]{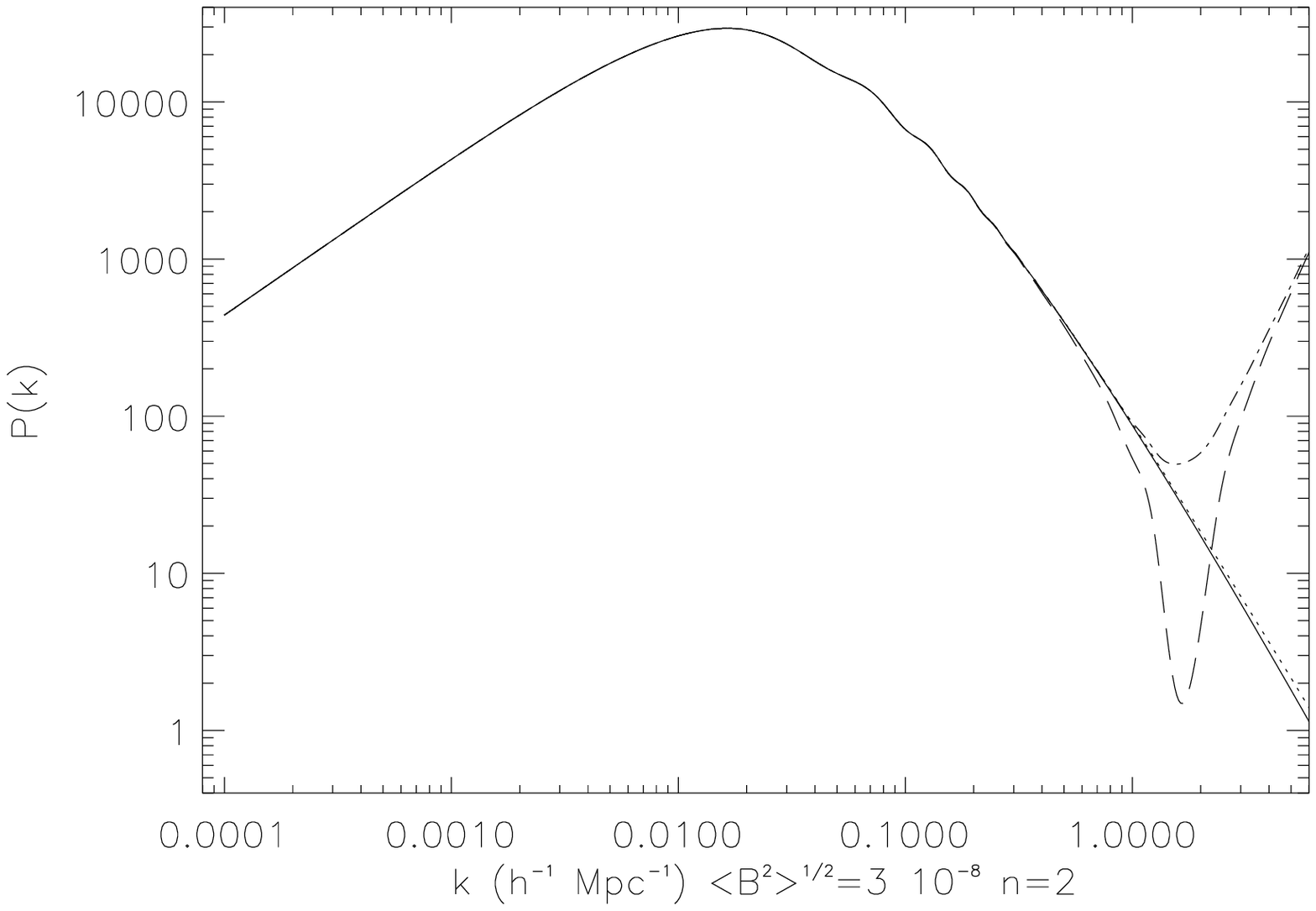}
\caption{Linear cold dark matter power spectrum obtained with fully correlated 
PMF with (dashed line) and without (dotted line) Lorentz term,
with uncorrelated PMF and the Lorentz force (dot-dashed line) in comparison 
with the vanishing PMF (solid line). In the figure
$\sqrt{\langle B^2 \rangle} = 3 \times 10^{-8} \,
{\rm Gauss}$, $k_D = 2 \pi$ and $n_B=2$ are considered. 
The other cosmological parameters are the same as Fig. (\ref{fig_thetab}).}
\label{matterps}
\end{figure}

In Fig. (\ref{matterps}) we present the results for the linear CDM power 
spectrum evaluated at present time in presence of SB of PMF. 
By analyzing Fourier spectra we have checked that the adiabatic results 
are recovered for $k > 2 k_D$. We compare the results obtained 
by neglecting or by taking into account the Lorentz term. 
By considering the equations evolved 
and the previous figures, it is clear how the Lorentz term treated as in 
Eq. (\ref{baryonseq}) is a leading contribution for baryons which gives 
rise to a long-time effect as show in Fig. 6. Through gravity  
CDM is affected as shown in Figs. (7-9) and therefore a large feature is 
present in the linear CDM matter PS.

\section{Conclusions}

We have investigated the impact of a SB of PMF on scalar 
cosmological perturbations and its impact on CMB anisotropies and matter 
power spectrum. The effects on the CMB angular power 
spectrum is one of the distinctive features of stochastic PMF together with 
non-gaussianities and Faraday rotation \cite{faraday}:
future missions as {\sc Planck} \cite{bluebook} will greatly 
improve the present constraints \cite{yamazaki,japan_constraints}. 

We have analyzed the SB of PMF in the one-fluid MHD 
approximation \cite{giovannini_review} as a source for cosmological 
perturbations and we have inserted such modifications in the CAMB code 
\cite{CAMB}.
Our numerical code improves previous studies \cite{yamazaki,giovanninikunze} 
for the treatment of initial 
conditions and exact convolutions for the PMF energy-momentum tensor. 
Note that the present constraints \cite{yamazaki,japan_constraints}
used neither the correct initial conditions nor the
correct convolutions for the PMF energy density and Lorentz force power
spectra. Ref. \cite{giovanninikunze} uses the correct initial conditions, 
but a power spectrum for the PMF energy -density 
with a spectral index which is twice the one for the power spectrum of the 
magnetic field. We have shown extensively in Sect. VI and 
Appendix A,B that this is not the case.

We have also shown how 
the Lorentz term for baryons in the one-fluid plasma description 
\cite{giovannini_review} may lead to a long-time 
effect which we have described analytically in Eq. (26). 
This last point deserves further investigation.

\acknowledgments

We are grateful to Chiara Caprini, Ruth Durrer and Jose Alberto 
Rubino-Martin for conversations and discussions on magnetic fields. 
This work has been done in the framework of the Planck LFI activities and 
is partially supported by 
ASI contract Planck LFI Activity of Phase E2.
We thank INFN IS PD51 for partial support. 
F. F. is partially supported by INFN IS BO 11.

\section{Note Added}

While this paper was about to be completed, an article 
\cite{japan_new} which computes numerically the convolution integrals 
by taking into account the angular part 
appeared. Ref. \cite{japan_new} does not display the dependence on 
$k$ of these convolution integrals and therefore we cannot compare 
our analytical expressions with their results.

\appendix

\section{Energy Density}

\begin{figure}
\includegraphics[scale=0.54]{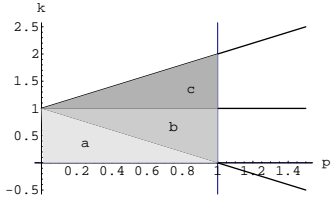}
\caption{Integration domains in $(k,p)$ plane}
\label{kpsplitting}
\end{figure}

The convolution which gives the magnetic energy density spectrum is
given in Eq. (\ref{edps}) with the parametrization for the PS for the magnetic 
field given in Eq. (\ref{bps}). Let us compute the energy density 
of magnetic field applying the sharp cut-off to both spectra in the 
convolution according to Eq. (\ref{spectrum2}). In this case we have
two conditions of existence to take into account:
\be
p<k_D\,, \qquad \qquad \qquad | \vec k-\vec p | <k_D
\ee
This second condition poses a $k$-dependence on the angular integration domain and, together with the first 
one, allows the energy power spectrum to be defined for $0<k<2k_D$. For simplicity of notation we normalize 
the Fourier wavenumber to $k_D$ and perform the integration with this choice. 
The double integral (over $\gamma$ and over $p$) we have to compute must therefore be splitted in three parts
depending on the $\gamma$ and $p$ lower and upper $k$-depending bounds. This splitting is well displayed
in $(k,p)$ plane as showed in Fig.(\ref{kpsplitting}): in region $a$ the angular integration has
to be done between $-1$ and $1$, while within $b$ and $c$ regions between $(k^2+p^2-1)/2kp$ and $1$. 

\begin{widetext}
A sketch of the integration is thus the following:
\begin{eqnarray}
1)&&  0<k<1 \nonumber\\
&&\int_{0}^{1-k}dp 
\int_{-1}^{1}d\gamma\,\dots + \int_{1-k}^{1}dp 
\int_{\frac{k^2+p^2-1}{2kp}}^{1}d\gamma\,\dots \equiv 
\int_{0}^{1-k}dp I_a (p,k) + \int_{1-k}^{1}dp I_b (p,k) 
\nonumber\\
2) && 1<k<2 \nonumber\\
&& \int_{k-1}^{1}dp 
\int_{\frac{k^2+p^2-1}{2kp}}^{1} d\gamma\,\dots \equiv 
\int_{k-1}^{1}dp I_c (p,k)
\label{intscheme}
\end{eqnarray}

The angular integrals can be performed as: 
\begin{eqnarray}
I_a&=&\int_{-1}^1 p^{n+2}\Bigg[2-
\frac{k^2(1-\gamma^2)}{k^2+p^2-2kp\gamma}\Bigg](k^2+p^2-2kp\gamma)^{n/2}d\gamma
\nonumber\\
&=&\frac{2p^{n-1}}{kn(2+n)(4+n)}\Big[(k+p)^{n+2}\Big(k^2-k(2+n)p
+(1+4n+n^2)p^2\Big)\nonumber\\
&&-|k-p|^{n+2}\Big(k^2+k(2+n)p+(1+4n+n^2)p^2\Big)\Big]\,,
\end{eqnarray}

\begin{eqnarray}
I_b=I_c&=&\!\!\!\int_{\frac{k^2+p^2-1}{2kp}}^1 p^{n+2}\Bigg[2-\frac{k^2(1-\gamma^2)}{k^2+p^2-2kp\gamma}\Bigg](k^2+p^2-2kp\gamma)^{n/2} d\gamma\nonumber\\
&=&\frac{p^{n-1}}{4kn(2+n)(4+n)}\Big[8k^4+2n-8k^2n+6k^4n+n^2-2k^2n^2+k^4n^2 \nonumber\\
&&-16k^2p^2+24np^2-12k^2np^2+6n^2p^2-2k^2n^2p^2+8p^4+6np^4+n^2p^4 \nonumber\\
&&-8| k-p|^{n+2}\big(k^2+k(2+n)p+(1+4n+n^2)p^2\Big)\Big]
\end{eqnarray}

\end{widetext}

Note that the divergent terms at the denominator $n$ and $n+2$ 
simply means that the above formulae are not applicable for $n=0$ and 
$n=-2$ (logarithmic terms appear in these cases).

A special care must be taken in the radial integral.In particular the presence of the term
 $|k-p|^{n+2}$ in both integrands, when considering odd spectral indexes, makes necessary to
 divide the two cases:$p<k$ and $p>k$.This leads to a further division of the integration domain.The scheme of the
radial integration is then:\\
\begin{displaymath}
\int_0^{(1-k)} dp\rightarrow
\left\{\begin{array}{ll}
  k<1/2
\left\{\begin{array}{ll}
\int_0^k dp...
\quad
& {\rm with}  \quad p < k  \\
\int_k^{(1-k)}dp...
& {\rm with}\quad p > k \\ 
\end{array} \right.\\
 k>1/2 \quad\int_0^{(1-k)}dp...\quad 
 {\rm with}\quad p < k \\
\end{array} \right. 
\end{displaymath}

\begin{displaymath}
\int_{(1-k)}^1 dp\rightarrow
\left\{\begin{array}{ll}
 k<1/2 \quad\int_{(1-k)}^1dp... \qquad
 {\rm with}\quad p > k \\ 
 k>1/2
\left\{\begin{array}{ll}
\int_{(1-k)}^k dp...
\quad
& {\rm with}  \quad p < k  \\
\int_k^1 dp...
& {\rm with}\quad p > k \\
\end{array} \right.
 \end{array} \right. 
\end{displaymath}

\begin{displaymath}
\int_{(k-1)}^1 dp\rightarrow
\left\{\begin{array}{ll}
 1<k<2 \quad\int_{(k-1)}^1dp... \quad
 {\rm with}\quad p < k \\
\end{array}  \right.
\end{displaymath}
\noindent
It is important to study some relevant behaviour of the integrands in $p$.
For $p \sim 0$: 
\be
I_a \sim \frac{8}{3} k^n p^{n+2} \,,
\label{I1psim0}
\ee

\begin{widetext}

For $p \sim k$ the above integrands behave as 
\be
I_a \sim \frac{2 k^{n-2}}{n (n+2) (n+4)} \left[  
2^{n+2} k^{n+4} n(n+3) 
- 
\left((k-p)^2\right)^{\frac{n+2}{2}} (n+1)(n+4) k^2 \right] \,,
\ee
\be
I_b \sim \frac{k^{n-2}}{4 n (n+2) (n+4)} 
\left[  n \left(4 (n+4) k^2 + n + 2 \right)  
- 8 \left((k-p)^2\right)^{\frac{n+2}{2}} (n+1)(n+4) k^2 \right] \,.
\ee
For $I_c$, $p \sim k$ cannot be obtained since $1 < k < 2$.
It is important to stress that for $n > -3$ the divergences in 
$p \sim 0$ and $p \sim k$ are integrable. The coefficients of both leading 
terms are proportional to $k^n$. 


Following the scheme (\ref{intscheme}) we can perform the integration over $p$.
Our exact results are given for particular values of $n_B$. 

\subsection{$n_B=4$}

\be
|\rho_B (k) |^2_{n_B=4} = \frac{A^2 k_D^{11}}{64\pi k_*^{8}}
\left[\frac{4}{11} - \tk + \frac{4}{3} \tk^2 - \tk^3 + 
\frac{8}{21} \tk^4 - \frac{\tk^5}{24} - \frac{\tk^7}{192}
+ \frac{\tk^{11}}{9856} \right]
\label{n4}
\ee


\subsection{$n_B=3$}

\begin{displaymath}
|\rho_B (k) |^2_{n_B=3} = 
\frac{A^2 k_D^{9}}{64\pi k_*^{6}}
\left\{\begin{array}{ll}
\frac{4}{9} - \tk + \frac{20}{21} \tk^2 
- \frac{5}{12} \tk^3 + \frac{4}{75} \tk^4 + \frac{4}{315} \tk^6 - 
\frac{\tk^9}{525} \quad
& {\rm for}  \quad 0 \le \tk \le 1 \\
(2-\tk)^2 \frac{264 - 436 \tk + 863 \tk^2 - 528 \tk^3 + 48 \tk^5 + 48 \tk^6 
+ 16 \tk^7 + 4 \tk^8}{6300 k}
& {\rm for}
\quad 1 \le \tk  \le 2
\end{array} 
\label{n3}
\right. 
\end{displaymath}

\subsection{$n_B=2$}

\be
|\rho_B (k) |^2_{n_B=2} = \frac{A^2 k_D^{7}}{64\pi k_*^{4}}
\left[ 
\frac{4}{7} - \tk + \frac{8}{15} \tk^2 - \frac{\tk^5}{24} 
+ \frac{11}{2240} \tk^7 \right]
\label{n2}
\ee


\subsection{$n_B=1$}

\begin{displaymath}
|\rho_B (k) |^2_{n_B=1} = 
\frac{A^2 k_D^5}{64\pi k_*^2}
\left\{\begin{array}{ll}
\frac{4}{5} - \tk + \frac{1}{4} \tk^3 
+ \frac{4}{15} \tk^4 - \frac{1}{5} \tk^5 \quad
& {\rm for}  \quad 0 \le \tk \le 1 \\
(2-\tk)^2 \frac{8 - 4 \tk - \tk^2 + 4 \tk^4}{60 k}
& {\rm for}
\quad 1 \le \tk  \le 2
\end{array} 
\label{n3}
\right. 
\end{displaymath}

\subsection{$n_B=0$}

\begin{displaymath}
|\rho_B (k) |^2_{n_B=0} = \frac{A^2 k_D^3}{64 \pi}
\left\{\begin{array}{ll}
\frac{1}{96 \tk} \left[ \tk (116 - 102 \tk - 84 \tk^2 + \tk^3 (53 + 4 \pi^2)) +
\right. \nonumber \\ 
\left. 
12 \log(1 - \tk) (-1 + 4 \tk^2 - 3 \tk^4 + 4 \tk^4 \log\tk ) - 
\right. \nonumber \\
\left. 48 \, \tk^4 \, {\rm PolyLog} [2,\frac{ -1 + \tk}{\tk}] \right] 
\quad
& {\rm for}  \quad 0 \le \tk \le 1 \\
\frac{1}{96 \tk} \left[ 116 \tk - 102 \tk^2 - 84 \tk^3 + 53 \tk^4 +
\right. \nonumber \\
\left. \log[-1 + \tk] (-12 + 48 \tk^2 - 36 \tk^4 + 24 \tk^4  \log\tk) + \right.
\nonumber \\   
\left. 24 \tk^4 {\rm PolyLog} [2, \frac{1}{\tk}] - 24 \tk^4 
{\rm PolyLog} [2, \frac{-1 + \tk}{\tk}] \right]
& {\rm for}
\quad 1 \le \tk \le 2
\end{array} 
\label{n0}
\right. 
\end{displaymath}

\subsection{$n_B=-1$}

\begin{displaymath}
|\rho_B (k) |^2_{n_B=-1} = \frac{A^2 k_D k_*^2}{64 \pi} 
\left\{\begin{array}{ll}
4 - 5 \tk + \frac{4 \tk^2}{3} +\frac{\tk^3}{4}
\quad 
& {\rm for}  \quad 0 \le \tk \le 1 \\ 
\frac{((-2 + \tk)^2 (8 - 4 \tk + 3 \tk^2))}{12 \tk}
& {\rm for}  
\quad 1 \le \tk  \le 2
\end{array} \right. 
\label{n-1}
\end{displaymath}

\subsection{$n_B=-3/2$}

\begin{displaymath}
|\rho_B (k) |^2_{n=-3/2} = \frac{A^2 k_*^3}{64 \pi} 
\left\{\begin{array}{ll}
\frac{1}{45} \left[ \frac{8 (-33 + 29 \tk - 4 \tk^2 + 8 \tk^3)}
{\sqrt{1 - \tk} \tk} 
+\frac{ 264}{\tk} + 60 \tk + 5 \tk^3 \right. \nonumber \\
\left. 
- 90 \pi + 360 \log[1 + \sqrt{1 - k}] - 180 \log \tk \right]
\quad & {\rm for}  \quad 0 \le \tk \le 1 \\ 
\frac{1}{45} \left[ -\frac{(8 (-33 + 29 \tk - 4 \tk^2 + 8 \tk^3)}
{\sqrt{-1 + \tk} \tk} + \frac{264}{\tk} + 
   60 k + 5 k^3 \right. \nonumber \\
\left. - 180 \arctan[\frac{1}{\sqrt{-1 + \tk}}] + 
180 \arctan[\sqrt{-1 + \tk}] \right] & {\rm for}  
\quad 1 \le \tk  \le 2
\end{array} \right. 
\label{n-32}
\end{displaymath}

\end{widetext}

\section{Lorentz Force}

In order to obtain the complete estimate of the contribution of PMFs to the 
perturbation evolution is necessary to solve the 
convolution for the Lorentz Force power spectrum.
The anisotropic stress can be obtained directly from its relation with the 
Lorentz force and the magnetic energy density.
The convolution which gives the Lorentz force is
given in Eq. (\ref{spectrum_LF}) with the parametrization for 
the PS for the magnetic
field given in Eq. (\ref{bps}).

\begin{widetext}

\subsection{$n_B=4$}

\begin{displaymath}
|L (k) |^2_{n_B=4} = \frac{A^2 k_D^{11}}{64 \pi k_*^8}
\left\{\begin{array}{ll}
\frac{4}{15} - \frac{16 \tk}{15} + \frac{317 \tk^2}{135} - \frac{10 \tk^3}{3} 
+ \frac{45 \tk^4}{14} - \frac{1481 \tk^5}{720} +\frac{4 \tk^6}{5} 
- \frac{977 \tk^7}{6720} + \frac{1357 \tk^{11}}{2661120}
\quad
& {\rm for}  \quad 0 \le \tk \le 1 \\
\frac{248}{1155} - \frac{2 \tk}{3} + \frac{137 \tk^2}{135} 
-\frac{5 \tk^3}{6} +\frac{5 \tk^4}{14} - 
\frac{41 \tk^5}{720} - \frac{17 \tk^7}{6720} + 
\frac{41 \tk^{11}}{532224} & {\rm for}
\quad 1 \le \tk  \le 2
\end{array} \right.
\end{displaymath}

\subsection{$n_B=3$}

\begin{displaymath}
|L (k) |^2_{n_B=3} = \frac{A^2 k_D^9}{64 \pi k_*^6}
\left\{\begin{array}{ll}
\frac{44}{135} - \frac{7 \tk}{6} + \frac{547 \tk^2}{245} -\frac{8}{3}\tk^3 + 
\frac{3293 \tk^4}{1575} - \tk^5+\frac{2363 \tk^6}{10395}-\frac{131\tk^9}{33075} 
\quad
& {\rm for}  \quad 0 \le \tk \le 1 \\
 \frac{1}{727650} \left(-344960 + \frac{1920}{\tk^5} - \frac{12320}{\tk^3} 
+ \frac{133056}{\tk} +582120 \tk - 
    585090 \tk^2 + 323400 \tk^3 + \right.\nonumber \\
\left.
- 66066 \tk^4 - 3710 \tk^6 + 319 \tk^9
\right)
& {\rm for}
\quad 1 \le \tk  \le 2
\end{array} \right. 
\end{displaymath}

\subsection{$n_B=2$}

\begin{displaymath}
|L (k) |^2_{n_B=2} = \frac{A^2 k_D^7}{64 \pi k_*^4}
\left\{\begin{array}{ll}
\frac{44}{105} - \frac{4 \tk}{3} + \frac{11 \tk^2}{5} -\frac{13\tk^3}{6} + 
\frac{4 \tk^4}{3} - \frac{33 \tk^5}{80} + \frac{7 \tk^7}{320} 
\quad
& {\rm for}  \quad 0 \le \tk \le 1 \\
 \frac{32}{105}- \frac{2 \tk}{3}+ \frac{3 \tk^2}{5}-\frac{\tk^3}{6} 
- \frac{ \tk^5}{80}+\frac{19 \tk^7}{6720}  
& {\rm for}
\quad 1 \le \tk  \le 2
\end{array} \right. 
\end{displaymath}

\subsection{$n_B=1$}

\begin{displaymath}
|L (k) |^2_{n=1} = \frac{A^2 k_D^5}{64 \pi k_*^2}
\left\{\begin{array}{ll}
\frac{44}{75} - \frac{5 \tk}{3} + \frac{761 \tk^2}{315} -2\tk^3 + 
\frac{659 \tk^4}{630} - \frac{41 \tk^5}{150} 
\quad
& {\rm for}  \quad 0 \le \tk \le 1 \\
 \frac{128 - 480 k^2 + 2240 k^4 - 4368 k^5 + 4200 k^6 - 1310 k^7 - 145 k^9 + 
 77 k^{10}}{3150 k^5}
& {\rm for}
\quad 1 \le \tk  \le 2
\end{array} \right. 
\end{displaymath}

\subsection{$n_B=0$}

\begin{displaymath}
|L (k) |^2_{n_B=0} = \frac{A^2 k_D^3}{64 \pi}
\left\{\begin{array}{ll}
\frac{1}{1152 \tk^5} \left[ 12 (-1 + \tk^2)^3 (3 + \tk^2) \log[1 - \tk] + \right. \nonumber \\ 
\left.   12 (-3 + 8 \tk^2 - 30 \tk^4 + 64 \tk^5 - 48 \tk^6 + 9 \tk^8) \log[1 - \tk] + \right. \nonumber \\
\left.   \tk (-72 - 36 \tk + 168 \tk^2 + 78 \tk^3 + 744 \tk^4 - 2484 \tk^5 + 4728 \tk^6 - 
2869 \tk^7 + 1536 \tk^7 \log \tk) \right]
\quad
& {\rm for}  \quad 0 \le \tk \le 1 \\
\frac{1}{1152 k^5} \left[(2 - k) k (-36 + 
k (-36 + k (66 + k (72 + k (152 + k (-14 + 53 k)))))) + \right. \nonumber \\
\left. 24 (-3 + k^2 (8 + k^2 (-18 + k (32 - 24 k + 5 k^3)))) \log(-1 + \tk)  \right]
& {\rm for}
\quad 1 \le \tk  \le 2
\end{array} \right. 
\end{displaymath}

\subsection{$n_B=-1$}

\begin{displaymath}
|L (k) |^2_{n_B=-1} = \frac{A^2 k_D k_*^2}{64 \pi}
\left\{\begin{array}{ll}
\frac{44}{15} - \frac{83 \tk^2}{35} + 4 \tk \log \tk
\quad
& {\rm for}  \quad 0 \le \tk \le 1 \\
\frac{-64 + \tk^2 (112 + \tk^3 (112 + \tk (-140 + 39 \tk)))}{105 \tk^5};
& {\rm for}
\quad 1 \le \tk  \le 2
\end{array} \right. 
\end{displaymath}

\subsection{$n_B=-3/2$}

\begin{displaymath}
|L (k) |^2_{n_B=-3/2} = \frac{A^2 k_*^3}{64 \pi}
\left\{\begin{array}{ll}
\frac{8}{9} - \frac{2048}{2925 \tk^5} + \frac{128}{135 \tk^3} + \frac{8}{5 \tk} 
- \frac{4 \tk}{3} - \frac{2 \pi}{15} + 
\frac {88}{15} \log[1 + \sqrt{1 - \tk}] \nonumber \\
- \frac{44 \log \tk}{15} - 
  \frac{8 (-768 + 384 \tk + 1136 \tk^2 - 472 \tk^3 + 1655 \tk^4 + 5455 \tk^5 - 
     10160 \tk^6 + 2770 \tk^7)}{8775 \sqrt{1 - \tk} \tk^5}
\quad
& {\rm for}  \quad 0 \le \tk \le 1 \\
\frac{4}{8775 \sqrt{-1 + \tk} \, \tk^5} 
\left[ -1536 - 1536 \sqrt{-1 + \tk} + 768 \tk + 2272 \tk^2 + \right. \nonumber \\
\left.  2080 \sqrt{-1 + \tk} \tk^2 - 944 \tk^3 + 3310 \tk^4 + 
3510\sqrt{-1 + \tk} \tk^4 - 4690 \tk^5 + \right. \nonumber \\
\left. 1950 \sqrt{-1 + \tk} \tk^5 - 820 \tk^6 - 2925\sqrt{-1 + \tk} \tk^6 + 1640 \tk^7 \right. \nonumber \\
\left. - 585 \sqrt{-1 + \tk} \tk^5 \arctan[\frac{1}{\sqrt{-1 + \tk}}] + 
585 \sqrt{-1 + \tk} \tk^5 \arctan[\sqrt{-1 + \tk}] \right]
& {\rm for}
\quad 1 \le \tk  \le 2
\end{array} \right. 
\end{displaymath}

\end{widetext}

\end{document}